\documentclass[10pt,journal,compsoc]{IEEEtran}
\ifCLASSOPTIONcompsoc
\usepackage[nocompress]{cite}
\else
\usepackage{cite}
\fi
\usepackage[pdftex]{graphicx}
\usepackage{amsthm}
\usepackage{algorithm}% http://ctan.org/pkg/algorithms
\usepackage{algpseudocode}% http://ctan.org/pkg/algorithmicx
\usepackage{array}
\usepackage{multirow}
\usepackage{tikz,pgfplots,pgfplotstable}
\usepackage{subfigure}
\usetikzlibrary{backgrounds,automata}
\usepackage{xcolor}
\usepackage{url}

\definecolor{filtering}{HTML}{D7191C}
\definecolor{verification}{HTML}{FDAE61}

\newenvironment{customlegend}[1][]{%
	\begingroup
	\csname pgfplots@init@cleared@structures\endcsname
	\pgfplotsset{#1}%
}{%
	\csname pgfplots@createlegend\endcsname
	\endgroup
}%

\def\addlegendimage{\csname pgfplots@addlegendimage\endcsname}
\newtheorem{definition}{Definition}

\begin{document}
	\title{Speeding-up the Verification Phase of Set Similarity Joins in the GPGPU paradigm}
	
	\author{Christos~Bellas,
		Anastasios~Gounaris~\IEEEmembership{Senior Member,~IEEE}
% <-this % stops a space
		\IEEEcompsocitemizethanks{\IEEEcompsocthanksitem C. Bellas and A. Gounaris are with the Aristotle University of Thessaloniki, Greece.\protect\\
Email: chribell,~gounaria@csd.auth.gr}% <-this % stops an unwanted space
	}

	\IEEEtitleabstractindextext{%
		\begin{abstract}
We investigate the problem of exact set similarity joins using a co-process CPU-GPU scheme. The state-of-the-art CPU solutions split the wok in two main phases. First, filtering and index building takes place to reduce the candidate sets to be compared as much as possible; then the pairs are compared to verify whether they should become part of the result. We investigate in-depth solutions for transferring the second, so-called verification phase, to the GPU addressing several challenges regarding the data serialization and layout, the thread management and the techniques to compare sets of tokens. Using real datasets, we provide concrete experimental proofs that our solutions have reached their maximum potential, since they totally overlap verification with CPU tasks, and manage to yield significant speed-ups, up to 2.6X in our cases.

		\end{abstract}
		
		}

	\maketitle

	\IEEEdisplaynontitleabstractindextext

	% \fi
	\IEEEpeerreviewmaketitle

	\IEEEraisesectionheading{\section{Introduction}\label{sec:introduction}}

	Given two collections of sets and a threshold, set similarity join is the operation of computing all pairs the overlap of which exceeds the given threshold. Similarity joins are used in a range of applications, such as plagiarism detection, web crawling, clustering and data mining and have been the subject of extensive research recently, e.g., \cite{mann:ssjoin,BayardoMS07,BourosGM12,JLFL14,SHC14}.
	
	In very large datasets, finding similar sets is not trivial. Due to the inherent quadratic complexity, a set similarity join between even medium sized datasets can take hours to complete on a single machine\footnote{E.g., a similarity join over the DBLP dataset using a Jaccard threshold of 0.85 takes 8.5 hours approx.}.  In addition, challenges like high dimensionality, sparsity, unknown data distribution and expensive evaluation arise. To tackle scalability challenges, two main and complementary approaches have been followed. Firstly, to devise sophisticated techniques, which safely prune pairs that cannot meet the threshold as early as possible, typically through simple computations related to the prefix and the suffix of the ordered sets, e.g. \cite{mann:ssjoin,JLFL14}. Secondly, to benefit from massive parallelism offered by the MapReduce framework, e.g., \cite{BML10,VCL10,MF12,SHC14}.
	
	In this work, we explore a third direction, namely to couple techniques of the first approach with the massive parallelism offered by modern graphics cards.
	Modern GPUs offer a high-parallel environment at low cost. As a result, general-purpose computing on graphics processing units (GPGPU) has been introduced \cite{KDK+11}. In general, GPGPU takes advantage of the different and complementary characteristics offered by CPUs and GPUs to improve performance. It has been employed in domains like deep learning, bioinformatics, numerical analytics and many others.
	However, implementing existing algorithms and techniques on the GPU requires in depth knowledge of the hardware and is often counter-intuitive. In addition, not all tasks are suitable for GPU-side processing. A traditional CPU surpasses at complex branching in application logic, while a GPU is superior at mass parallel execution of simple tasks and floating point operations \cite{mittal2015survey}.
	
	To date, the problem of exact set similarity on GPUs has not been addressed apart from \cite{Ribeiro-JuniorQ17}; however, the proposal in \cite{Ribeiro-JuniorQ17} does not go into the implementation details, which is our focus. In general, in the literature, there exist several proposals for approximate set similarity or for similar problems, such as nearest neighbor search, e.g., \cite{lieberman:gpusjoin,cruz2015gpu,johnson2017billion} (the detailed discussion of related work is deferred to Section \ref{sec:rw}). Therefore, there is a gap in detailed investigation of exact similarity joins in GPUs.
	The main contribution of this work is to fill this gap and propose efficient solutions after thoroughly investigating several design alternatives.
	%The research to date has tended to focus on putting the whole workload on one side while the other remains idle.
	
Our work incorporates a co-process scheme between CPU and GPU in order to efficiently compute set similarity join. In our scheme, CPU remains responsible for index building and initial pruning of candidate pairs, whereas the GPU computes the overlap of all remaining pairs. This  however leads to limited maximum speed-ups because of the Amdahl's law. Moreover, the GPU part comes with
several challenges regarding the data serialization and layout, the thread management and the techniques to compare sets of tokens. 
In this work, we address all challenges and show that we manage to achieve speed-ups up to 2.6X;  moreover, we show that our scheme has reached its maximum potential in the sense that it annihilates the impact of GPU tasks on the running time.
	%Finally, we capitalize on the results of the comparative study in \cite{mann:ssjoin}, which identifies the most efficient algorithms in a main-memory CPU setting, and we focus on transferring these algorithms in a CPU-GPU co-processing scheme.
	%It is heavily influenced by the work of Mann et al \cite{mann:ssjoin}.
		In summary, the technical contributions of our work are as follows:
	
	\begin{itemize}
		\item We provide a detailed implementation analysis and we propose alternatives that differ in the workload allocated to each GPU thread. We use the CUDA programming model, which is proprietary to NVIDIA \cite{cuda-book} but widespread in practice. \footnote{the code is publicly available from \url{https://github.com/chribell/gpussjoin} }
		\item We conduct extensive performance analysis on seven real world datasets. We compare our findings to the state-of-art CPU implementations and point out a number of optimizations to further increase the performance.
		\item We provide evidence that in settings where the candidate pairs are tens of billions or more, our solutions reach their maximum potential, since they manage to fully hide the impact of GPU tasks on the running time due to overlapped execution with the CPU.
	\end{itemize}
	
	\emph{Paper outline.} Next, we  provide  details on CUDA programming model and the main CPU-based approaches to set similarity joins. We present our solutions and the design alternatives involved in Section~\ref{sec:approach}. Technical details  are in Section~\ref{sec:impl}.
	The  experimental results are in Section~\ref{sec:eval} and the related work in Section~\ref{sec:rw}. Finally, we conclude in Section~\ref{sec:concl}.
	
	\section{Background}
	\label{sec:back}
	
	We provide a comprehensive overview of the CUDA programming model and explain its main concepts.
	We also introduce the filter-verification framework used by state-of-the-art main memory set similarity join algorithms in line with the comparison work conducted by Mann et al \cite{mann:ssjoin}.
	%In this, various set similarity join algorithms  are evaluated through a CPU-only implemented framework.
	
	\subsection{CUDA Overview}
	
	In CUDA terminology, CPU and the main memory are referred to as \textit{host}, while the GPU and its own memory are referred to as \textit{device}. In this work, we use the terms CPU and host (resp. GPU and device) interchangeably.

	\textit{Architecture.} CUDA-enabled GPUs have many cores called \textit{Streaming Processors (SPs)}, which are divided into groups called \textit{Streaming Multiprocessors (SMs)}. Each SM includes also other units such as ALUs, instruction units, memory caches for load/store operations, and follows the \textit{Single Instruction Multiple Data (SIMD)} parallel processing paradigm.
	
	\textit{Thread Organization.} Threads are organized in logical blocks called \textit{thread blocks}. A thread block is scheduled and executed in its entirety on a SM in groups of 32 threads called \textit{warps}. Threads within a warp are called \textit{lanes} and share the same instruction counter, thus they are executed simultaneously in a SIMD manner. There are two cases of thread divergence, which degrade performance, namely \textit{inter-warp}, when concurrent warps run unevenly and \textit{intra-warp}, when warp lanes take different execution paths. The latter is also simply referred to as \textit{warp divergence}.
		
	\textit{Memory hierarchy.} There are several memory types on CUDA-enabled GPUs. They are divided into on-chip and off-chip ones. Off-chip memories include the \textit{global}, \textit{constant}, \textit{texture} and \textit{local} memory. The global memory is the largest (in the order of GBs) but slowest memory. Data transferred from the host to device resides in global memory and it is visible to all threads. The constant memory is read-only and  much smaller (in the order of KBs). It is used for short access times on immutable data throughout the execution. The texture memory is essentially a read-only global memory  and is preferred when 2-dimensional spatial locality occurs in memory access patterns. The local memory is part of the global memory and is used when the registers needed for a thread are fully occupied or cannot hold the required data. This is called \textit{register spilling}. On-chip memories include the \textit{caches}, the \textit{shared} memory and the \textit{registers}. For data reuse, there are caches per SM and the L2 cache is shared across all SMs. The shared memory is the second fastest memory type. Each SM has its own shared memory. Data stored in shared memory can be accessed by all threads within the same block, thus threads of the same block are allowed to inter-communicate via shared-memory. The registers are the fastest memory type and  contain the instructions of a single thread and the local variables during the lifetime of that thread.
	
	\textit{Kernel grid.} Every function to be processed in parallel by the GPU is called a \textit{kernel}.  Each kernel is executed by multiple thread blocks, which form the kernel \textit{grid}. The grid can be regarded as an array of blocks with up to two dimensions.  Each block is, in turn, an array of threads with up to three dimensions. CUDA can schedule blocks to run concurrently on a SM depending on the shared memory and registers used per block. Increasing either of these factors can lead to limited  concurrent block execution, which results in low occupancy. \textit{Occupancy} is defined as the ratio of active warps on a SM to the maximum allowed active warps per SM. Maximizing occupancy is a good heuristic approach but it does not always guarantee performance gain. On the contrary, maintaining high warp \textit{execution efficiency}, i.e. the average percentage of active threads in each executed warp, is a more robust approach for data-management tasks, as shown in executing generic theta-joins on GPUs \cite{BellasG17}.
	
	\begin{figure}[tb!]
		\centering
		\includegraphics[width=0.65	\linewidth]{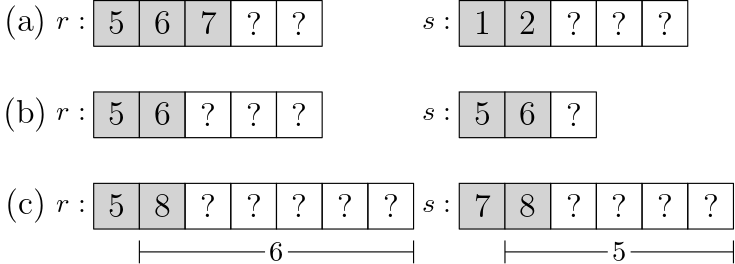}
		\caption{Filters used for candidate pruning: (a) prefix, (b) length, (c) positional}
		\label{fig:filters}
	\end{figure}

	\subsection{Set Similarity joins}
	
	The state-of-the-art main memory set-similarity algorithms conform to a common filter-verification framework, as explained in \cite{mann:ssjoin}, in which seven key representatives\footnote{AllPairs \cite{BayardoMS07}, PPJoin and PPJoin+ \cite{XiaoWLYW11}, MPJoin \cite{RIBEIRO201162}, MPJoin-PEL \cite{pel}, AdaptJoin \cite{WangLF12} and GroupJoin \cite{BourosGM12}.} are compared using real world datasets. The common idea behind all these algorithms is (i) to avoid comparing all possible set pairs by applying filtering techniques on preprocessed data to prune as much candidate pairs as possible; and (ii) then to proceed to the actual verification of the remaining candidates.
	We summarize the key points of the work of \cite{mann:ssjoin} that are relevant to our research below.
	
	\subsubsection{Data layout.} Every dataset is a collection of multiple sets. Each set consists of elements called \textit{tokens}. The data preprocessing phase involves a tokenization technique and deduplication of tokens. % using a counter that is appended on each duplicate, thus increasing the number of overall tokens.
	As a result of such a preprocessing phase, all the tokens of a set are unique.
	The input data tokens are represented by integers and are sorted by their frequency in increasing order, so that  infrequent tokens appear first in a set.
	The sets of a collection are sorted first by their size and then lexicographically within each block of sets of equal size.
	
	\subsubsection{Set Similarity functions.} To measure the similarity between sets, Jaccard, Dice and Cosine normalized similarity functions are typically used. The given normalized threshold $t_{n}$ is translated to an equivalent overlap $t$, which defines the minimum number of tokens that need to be shared between two sets to satisfy the threshold (see Table \ref{tab:sim_funcs}). For example, if the Jaccard similarity threshold of two 10-token sets is set to 0.8, this is translated to an overlap threshold of 9 tokens that need to be shared.
	\begin{table}[tb!]
		\caption{Similarity Functions (adapted from \cite{mann:ssjoin})}
		\label{tab:sim_funcs}
		\centering
		\begin{tabular}{c||c||c}
			Similarity function & Definition & Equivalent Overlap\\
			\hline
			Jaccard				& $\frac{|r \cap s|}{|r \cup s|}$  	& $\lceil\frac{t_{n}}{1+t_{n}}(|r|+|s|)\rceil$ \\
			&&\\
			Cosine				& $\frac{|r \cap s|}{\sqrt{|r||s|}}$& $\lceil t_{n}\sqrt{|r||s|}\rceil$ \\
			&&\\
			Dice				& $\frac{2|r \cap s|}{|r|+|s|}$		& $\lceil \frac{t_{n}(|r|+|s|)}{2}\rceil$ \\
			&&\\
			Overlap				& $|r \cap s|$			  			& $t$ \\
		\end{tabular}
	\end{table}

	\subsubsection{Filters.} The most widely used filter, called \textit{prefix-filter}, exploits the given threshold and similarity function by examining only two subsets, one from each set in the candidate pair, and discards the pair if there is no overlap between the subsets. For example, in Figure~\ref{fig:filters}(a), there is no overlap between the respective set prefixes, thus, even if there is an overlap on the remaining tokens,
any overlap threshold set to 4 or higher cannot be reached, and in such cases, the candidate pair $(r,s)$ can be safely pruned.

Another filter, noted as \textit{length filter}, takes advantage of the normalized similarity functions dependency on set size. Hence, a candidate pair can be pruned if the set size inequality $t_{n} \cdot |r| \leq |s| \leq |r|/t_{n} $ is not satisfied. In Figure~\ref{fig:filters}(b), if $t_{n}=0.8$, the shown candidate pair $(r,s)$ can be pruned despite the prefix equality because a 6-token $r$ set requires a $s$ set of size $4 \leq |s| \leq 6$.

The last filter used in the examined algorithms is the \textit{positional filter}. Given the first match position, it evaluates if a candidate pair can reach the similarity threshold. As an example, in Figure~\ref{fig:filters}(c), if the threshold implies that at least 6 tokens should be shared, the pair is pruned since the remaining tokens from set $s$ are not enough to reach the similarity threshold.

	\subsubsection{Algorithm outline.} The set similarity join operation is achieved by executing an index nested loop join consisting of three steps. First, through an index lookup and a length filter application, a preliminary candidate set (pre-candidates) is generated. In the second step, pre-candidates are deduplicated and filtered. The pairs that pass all filters form the final candidates. These two steps compose the \textit{filtering} phase. In the third and final step, also noted as \textit{verification} in the literature, the similarity score for each of the remaining candidate pair is computed and if it exceeds  the threshold, the pair is added to the output result.
	
	%\begin{algorithm}
	%	\caption{Set similarity join nested loop}\label{euclid}
	%	\begin{algorithmic}[1]
	%		\State Precandidates
	%		\State Prune further precandidates
	%		\State Verify final candidate pairs
	%	\end{algorithmic}
	%\end{algorithm}
	
	A special case is where the set similarity join is a self-join using only a single collection of sets. In that case, a token set is first probed against the current index contents and then added to the index itself. This allows for incremental index building that is interleaved with verification. Also, the fact that a set that probes the index is always no shorter than the current indexed sets can be leveraged to speed-up verifications.

	\begin{table}[!t]
		\renewcommand{\arraystretch}{1.1}
		\caption{Notation}
		\label{tab:notation}
		\centering
		\begin{tabular}{c||c}
			\hline
			$R$, $S$ & Collections of sets  to be joined \\
			\hline	
			$r_{i}$ (resp.$s_j $)  & a token set from $R$ (resp. $S$)\\		\hline			
			$t$		 & Similarity threshold \\		\hline	
			$C \subseteq R \times S $		 & Set of candidate pairs \\	\hline
			$O$		 & Device output \\	\hline	
			$R_{T}$, $S_{T}$ & Token arrays \\		\hline
			$R_{O}$, $S_{O}$, $C_{O}$ & Offset arrays \\		\hline
			$||R_{T}||$, $||S_{T}||$,$||C||$,$||O||$ & \multirow{ 2}{*}{Size of arrays in bytes} \\
			$||R_{O}||$, $||S_{O}||$,$||C_{O}||$&\\			\hline
			$H_{i}$  & The $i^{th}$ host thread, $i \in {0,1,2} $ \\		\hline
			$T$		 & Number of device threads \\		\hline
			$B$		 & Thread block size \\		\hline
			$M_{h}$	 & Host memory \\		\hline
			$M_{d}$	 & Device memory \\			\hline
			$M_{c} <  M_{d}$	 & Device memory for candidate pairs\\			\hline
		\end{tabular}
	\end{table}

	\section{Our Approach}
	\label{sec:approach}
	
	Let $R$ and $S$ be two collections of token sets. The set-similarity problem assumes that there exist a similarity function  $Similarity()$ and a user-defined threshold $t \in [0,1]$. It is formally defined as follows.
	
	\begin{definition}
		Problem Definition of set similarity joins: Find all pairs ($r,s), r \in R,~s \in S$ such that $Similarity(r,s) \ge t$.
	\end{definition}
	
	In a naive solution, the set of candidate pairs $C$ to be checked in the verification phase is all pairs $R \times S $. However, due to the filtering phase, for thresholds not close to 0, $C$ is typically a small subset of the cartesian product. The output of the verification is denoted as $O$.
	
	In our solution, we assume that the host (resp. device) is equipped with $M_h$ (resp. $M_d$) memory capacity. The host runs 3 threads ($H_0,~H_1,~H_2$), while the device executes $T$ threads in blocks of size $B$.  The collections of token sets are transferred in the device main memory in a linearized form, denoted as $R_{T}$ and $S_{T}$, and are accompanied by offset arrays $R_{O}$ and $S_{O}$ in order to distinguish the set boundaries.
	Table \ref{tab:notation} summarizes the notation.
	%%%%%%%%%%%%%%%%%%
	\pgfplotstableread{
		name
		ALL
		PPJ
		GRP
		ALL
		PPJ
		GRP
		ALL
		PPJ
		GRP
		ALL
		PPJ
		GRP
		ALL
		PPJ
		GRP
		ALL
		PPJ
		GRP
	}\datatable
	
	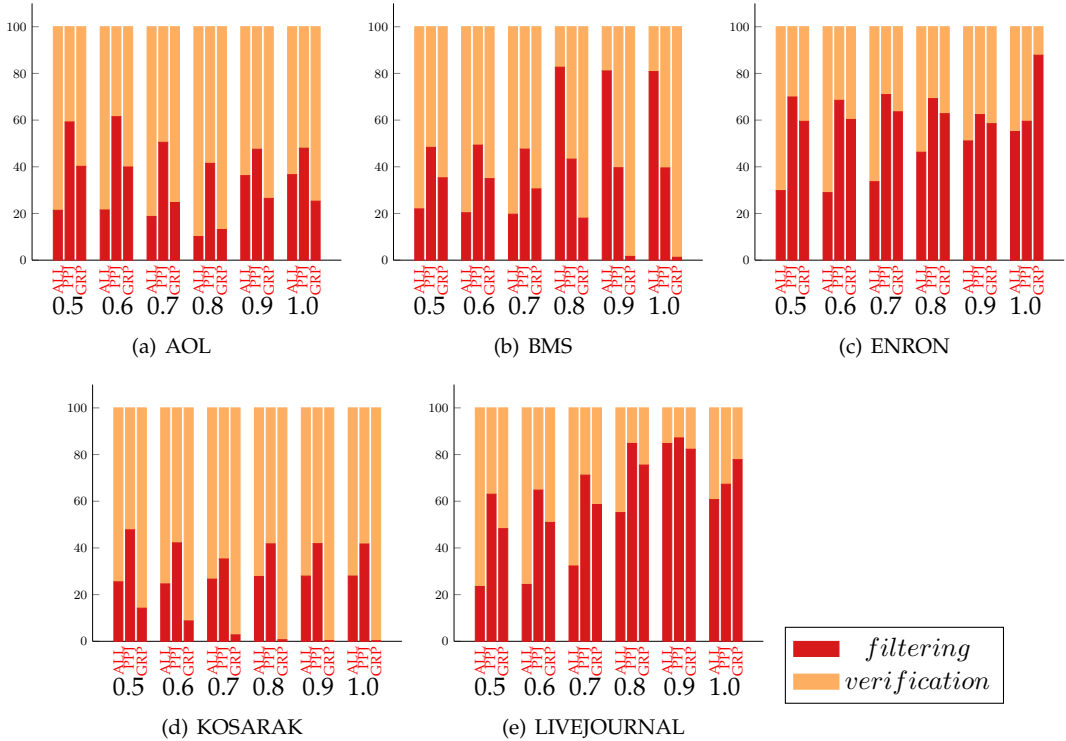
\begin{figure*}[t]
		\centering
		
		\subfigure[AOL]
		{
			\begin{tikzpicture}[scale = 0.6]
			
			\begin{axis}[ybar stacked,
			legend style={legend columns=3,
				at={(xticklabel cs:0.5)},
				anchor=north,
				draw=none},
			xtick=data,
			bar width=2mm,
			ymin=0,
			axis y line*=none,
			axis x line*=none,
			xticklabels from table={\datatable}{name},
			x tick label style={rotate=90,anchor=east,font=\small,color=red},
			tick label style={font=\footnotesize},
			legend style={font=\footnotesize,yshift=-3ex},
			label style={font=\footnotesize},
			xlabel style={yshift=-3ex},
			area legend]
			\addplot [filtering,fill=filtering,x tick label style={xshift=-0.3cm}] table[x=Count,y=filtering] {cpu_aol.txt};
			\addplot [verification,fill=verification,x tick label style={xshift=-0.3cm}] table[x=Count,y=verification] {cpu_aol.txt};
			
			\end{axis}
			\draw (0.8,-1.0) node{0.5};
			\draw (1.9,-1.0) node{0.6}; %+1.25
			\draw (2.9,-1.0) node{0.7}; %+1.30
			\draw (3.9,-1.0) node{0.8}; %+1.25
			\draw (5.0,-1.0) node{0.9}; %+1.30
			\draw (6.0,-1.0) node{1.0}; %+1.25
			\end{tikzpicture}
		}
		\subfigure[BMS]
		{
			\begin{tikzpicture}[scale = 0.6]
			
			\begin{axis}[ybar stacked,
			legend style={legend columns=3,
				at={(xticklabel cs:0.5)},
				anchor=north,
				draw=none},
			xtick=data,
			bar width=2mm,
			ymin=0,
			axis y line*=none,
			axis x line*=none,
			xticklabels from table={\datatable}{name},
			x tick label style={rotate=90,anchor=east,font=\small,color=red},
			tick label style={font=\footnotesize},
			legend style={font=\footnotesize,yshift=-3ex},
			label style={font=\footnotesize},
			xlabel style={yshift=-5ex},
			area legend]
			\addplot [filtering,fill=filtering,x tick label style={xshift=-0.3cm}] table[x=Count,y=filtering] {cpu_bms.txt};
			\addplot [verification,fill=verification,x tick label style={xshift=-0.3cm}] table[x=Count,y=verification] {cpu_bms.txt};
			
			\end{axis}
			\draw (0.8,-1.0) node{0.5};
			\draw (1.9,-1.0) node{0.6}; %+1.25
			\draw (2.9,-1.0) node{0.7}; %+1.30
			\draw (3.9,-1.0) node{0.8}; %+1.25
			\draw (5.0,-1.0) node{0.9}; %+1.30
			\draw (6.0,-1.0) node{1.0}; %+1.25
			\end{tikzpicture}
		}
		\subfigure[ENRON]
		{
			\begin{tikzpicture}[scale = 0.6]
			
			\begin{axis}[ybar stacked,
			legend style={legend columns=3,
				at={(xticklabel cs:0.5)},
				anchor=north,
				draw=none},
			xtick=data,
			bar width=2mm,
			ymin=0,
			axis y line*=none,
			axis x line*=none,
			xticklabels from table={\datatable}{name},
			x tick label style={rotate=90,anchor=east,font=\small,color=red},
			tick label style={font=\footnotesize},
			legend style={font=\footnotesize,yshift=-3ex},
			label style={font=\footnotesize},
			xlabel style={yshift=-5ex},
			area legend]
			\addplot [filtering,fill=filtering,x tick label style={xshift=-0.3cm}] table[x=Count,y=filtering] {cpu_enron.txt};
			\addplot [verification,fill=verification,x tick label style={xshift=-0.3cm}] table[x=Count,y=verification] {cpu_enron.txt};
			
			\end{axis}
			\draw (0.8,-1.0) node{0.5};
			\draw (1.9,-1.0) node{0.6}; %+1.25
			\draw (2.9,-1.0) node{0.7}; %+1.30
			\draw (3.9,-1.0) node{0.8}; %+1.25
			\draw (5.0,-1.0) node{0.9}; %+1.30
			\draw (6.0,-1.0) node{1.0}; %+1.25
			\end{tikzpicture}
		}
		\subfigure[KOSARAK]
		{
			\begin{tikzpicture}[scale = 0.6]
			
			\begin{axis}[ybar stacked,
			legend style={legend columns=3,
				at={(xticklabel cs:0.5)},
				anchor=north,
				draw=none},
			xtick=data,
			bar width=2mm,
			ymin=0,
			axis y line*=none,
			axis x line*=none,
			xticklabels from table={\datatable}{name},
			x tick label style={rotate=90,anchor=east,font=\small,color=red},
			tick label style={font=\footnotesize},
			legend style={font=\footnotesize,yshift=-3ex},
			label style={font=\footnotesize},
			xlabel style={yshift=-5ex},
			area legend]
			\addplot [filtering,fill=filtering,x tick label style={xshift=-0.3cm}] table[x=Count,y=filtering] {cpu_kosarak.txt};
			\addplot [verification,fill=verification,x tick label style={xshift=-0.3cm}] table[x=Count,y=verification] {cpu_kosarak.txt};
			
			\end{axis}
			\draw (0.8,-1.0) node{0.5};
			\draw (1.9,-1.0) node{0.6}; %+1.25
			\draw (2.9,-1.0) node{0.7}; %+1.30
			\draw (3.9,-1.0) node{0.8}; %+1.25
			\draw (5.0,-1.0) node{0.9}; %+1.30
			\draw (6.0,-1.0) node{1.0}; %+1.25
			\end{tikzpicture}
		}
		\subfigure[LIVEJOURNAL]
		{
			\begin{tikzpicture}[scale = 0.6]
			
			\begin{axis}[ybar stacked,
			legend style={legend columns=3,
				at={(xticklabel cs:0.5)},
				anchor=north,
				draw=none},
			xtick=data,
			bar width=2mm,
			ymin=0,
			axis y line*=none,
			axis x line*=none,
			xticklabels from table={\datatable}{name},
			x tick label style={rotate=90,anchor=east,font=\small,color=red},
			tick label style={font=\footnotesize},
			legend style={font=\footnotesize,yshift=-3ex},
			label style={font=\footnotesize},
			xlabel style={yshift=-5ex},
			area legend]
			\addplot [filtering,fill=filtering,x tick label style={xshift=-0.3cm}] table[x=Count,y=filtering] {cpu_livejournal.txt};
			\addplot [verification,fill=verification,x tick label style={xshift=-0.3cm}] table[x=Count,y=verification] {cpu_livejournal.txt};
			\end{axis}
			\draw (0.8,-1.0) node{0.5};
			\draw (1.9,-1.0) node{0.6}; %+1.25
			\draw (2.9,-1.0) node{0.7}; %+1.30
			\draw (3.9,-1.0) node{0.8}; %+1.25
			\draw (5.0,-1.0) node{0.9}; %+1.30
			\draw (6.0,-1.0) node{1.0}; %+1.25
			
			\end{tikzpicture}
		}
		\subfigure
		{
			\begin{tikzpicture}
			\begin{customlegend}[legend entries={$filtering$,$verification$}]
			\addlegendimage{filtering,fill=filtering,area legend}
			\addlegendimage{verification,fill=verification,area legend}
			\end{customlegend}
			\end{tikzpicture}
		}
		\caption{Filtering - Verification contribution to the total execution time for AllPairs (ALL), PPJoin (PPJ) and  GroupJoin(GRP) and thresolds $0.5 \le t \le 1$}
		\label{fig:cpu_percentages}
	\end{figure*}
	%%%%%%%%%%%%%%%%%%
	
	\subsection{Main Rationale}
	
	As already explained, the algorithms solving the set similarity problem efficiently conform to the filter-verification framework.
	The filtering phase involves probing index structures such as a hashtable. Although there has been some work on implementing hashtables and inverted lists on GPUs \cite{ashkiani2017dynamic}, which can be employed in exact set similarity joins as in \cite{Ribeiro-JuniorQ17}, we choose this phase to remain a CPU task. On the other hand, the verification phase is more suitable for parallelization, as it involves a merge-like loop, where the overlap of candidate pairs is computed. As soon as the overlap threshold is met or cannot be reached, the operation terminates. True positives must be verified and the necessary overlap is still computed, while the rejection of pairs in this stage leads to less token comparisons without sacrificing accuracy. Even though the average verification runtime is reported as constant for most datasets, employing the GPU for this part with a view to improving the overall performance is the main goal of this work.
	
	In summary, in our approach, we allocate initial indexing and filtering to the CPU and the verification phase to the GPU. Next, we exploit the results of \cite{mann:ssjoin}, according to which the three algorithms in the skyline, i.e., the most efficient ones are AllPairs, PPJoin and GroupJoin. We experiment with 5 real datasets on a NVIDIA Titan Xp GPU (full details are provided in Section \ref{sec:eval}). In Figure~\ref{fig:cpu_percentages}, we demonstrate the percentage of each phase on the total execution time. More specifically, due to the pipelined nature of processing, we present the \textit{upper bound} of the verification fraction and the  \textit{lower bound} of the filtering one.
The key observation of our tests is that indexing and filtering contributes to the total running time significantly. Therefore, due to the Amdahl's law, employing the GPU for the verification in a ideal setting is expected to yield improvements of several times, but lower of an order of magnitude. Our techniques manage to achieve this (e.g., see Figure \ref{fig:scalability}).
	
	Based on the results both from  \cite{mann:ssjoin} and our tests, in the following, we focus solely on AllPairs, PPJoin and GroupJoin.
%AllPairs is the simplest algorithm that applies a prefix and length filter \cite{BayardoMS07}. GroupJoin applies these two filters after grouping sets according to the common prefix, and also employs a positional filter \cite{BourosGM12}.
The key characteristics of the techniques are summarized below.
	
	\noindent\textbf{AllPairs (ALL)}.  It is the first and most naive main memory algorithm to exploit the given threshold. During the inverted list lookup, it applies the prefix and length filters to prune candidate pairs \cite{BayardoMS07}.
	
	\noindent
	\textbf{PPJoin (PPJ)}. It extends ALL by applying the positional filter on pre-candidates \cite{XiaoWLYW11}; therefore its verification phase is less loaded at the expense of a higher overhead during filtering.
	
	\noindent
	\textbf{GroupJoin (GRP)}. It is an extension to PPJ. Sets with identical prefix are \textit{grouped} together. Each group is handled as a single set. Thus GRP has faster indexing, as it discards candidates pairs in batches. During the verification phase, the candidate pairs are expanded \cite{BourosGM12}.

	When looking for similar sets in a single collection, which is the most common case, instead of looking for similar pairs between two different collections,
	the whole join process can be done incrementally, i.e. for a probing set, first its candidates are verified and then the algorithm proceeds to the next set. Naively
	allocating and copying small chunks of data on the GPU through a different kernel invocation per probing set, would incur an enormous overhead penalty. A more efficient alternative is to copy a large chunk of data stored in linear memory space to the GPU, process it there and copy back the results. Adopting this approach also improves overall runtime as the CPU builds candidate pair collections in waves and feeds them in a non-blocking manner to the GPU, which conducts the verification. Thus, time overlapping between the CPU and GPU tasks can be achieved. More specifically, we propose a multithreaded framework regarding both the host and device tasks. We thoroughly analyze each side below.
	
	Before delving into the details, we need to state our limitations. The most dominant constraining factor is the limited GPU memory. Due to this, the workload must be divided into chunks and the GPU should be invoked several times. The memory limitation mostly relates to the output size.
	As we have no prior knowledge about the output size, to ensure correctness the most straightforward solution is to allocate enough memory for the worst case, which is $O(|R||S|)$.
	
	\subsection{Host Tasks}
	
\begin{figure}[tb!]
		\centering
		\includegraphics[width=0.6\linewidth]{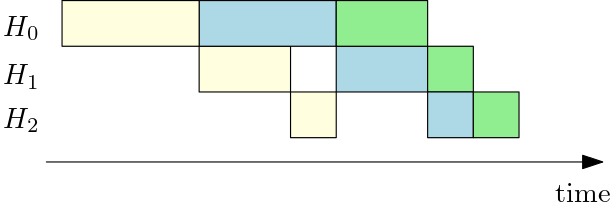}
		\caption{Execution overlap between host ($H_0,H_2$) and device ($H_1$)}
		\label{fig:executionoverlap}
	\end{figure}

	The host side is responsible for the filtering phase and works as the coordinator. Specifically, the host runs three threads. The first thread, $H_{0}$, conducts any filtering and builds chunks of candidates. When each chunk is built, the second thread, $H_{1}$, noted as \textit{device handler}, enacts the verification phase by copying the chunk to device memory and launching the kernel code. Meanwhile, $H_{0}$ continues to build the next chunk of candidates. As soon as the device output is copied back to host memory, the third thread $H_{2}$ post-process it to form the final pairs result. Note that $H_{2}$ may not be invoked if an aggregation is performed on top of the join, i.e., if only the count of pairs is needed instead of the actual pairs. In such a case, the device counts the number of pairs and returns the result to $H_{1}$. A visual representation of the execution overlap is depicted in Figure~\ref{fig:executionoverlap}, where each color corresponds to a different data chunk.
	
	\subsection{Device Tasks}
	
	The device side is responsible for the verification phase. It is invoked when the host prepares a chunk of candidate pairs. We present our data layout approach and discuss its impact on device memory.
	
	There are two levels of concurrency in CUDA, \textit{grid} and \textit{kernel}. Grid level concurrency concerns mostly the overlap between computation and data transfers, while kernel level concurrency refers to how a single task is executed in parallel by many threads \cite{cheng2014professional}.
	
	On the grid level, we further divide each input chunk of candidates into smaller chunks, each assigned to a different block. Thus we enhance the  overlapping between device computation and host-to-device data transfer. On the kernel level, we summarize our approaches as high-level verification alternatives (described in detail below).
	
	\subsubsection{Data Layout}
	
	By default, we pass data to device as arrays stored in consecutive memory space. This is preferred because parallel execution benefits from coalesced global memory accesses. However, due to the nature of the problem,  divergence in global memory access patterns is unavoidable, therefore the exploitation of on-chip memories is required to alleviate performance bottlenecks.
	\begin{figure}[tb!]
		\centering
		\includegraphics[width=0.85	\linewidth]{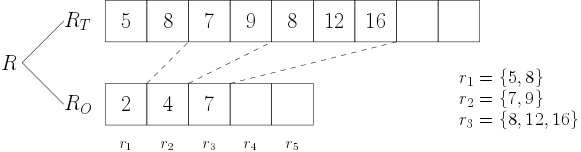}
		\caption{Example layout of a collection of threes sets in the device memory}
		\label{fig:datasetslayout}
	\end{figure}
	
	According to the linear memory layout, a collection $R$ is physically implemented as the composition of two arrays: tokens $R_{T}$ and offsets $R_{O}$. The former holds every token of every set in the collection in a sequence, while the latter is used to delimit each set boundaries. Figure~\ref{fig:datasetslayout} depicts how a collection of sets $R$ is stored in the device memory. Collection $S$ is stored in similar fashion. We transfer any collection of sets once in the beginning of the process.

	\begin{figure}[tb!]
		\centering
		\includegraphics[width=0.85	\linewidth]{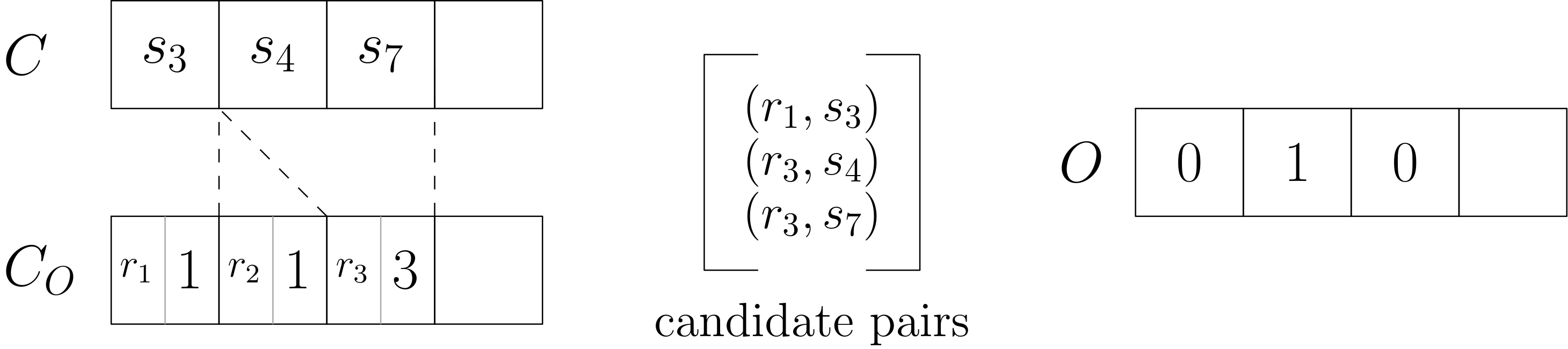}
		\caption{Candidates and output layout in device memory}
		\label{fig:candidatesoutputlayout}
	\end{figure}
	
	When the device is invoked to perform the verification phase, the host transfers an array of set IDs, noted as $C$, alongside with an array of offsets ($C_{O}$) which indicates the candidate pairs to be evaluated. In addition, an array of equal length to $C$, noted as $O$, is allocated on the device and it is used to store the output result. Essentially, $O$ is an array of boolean flags where true indicates that the corresponding candidate pair similarity is equal or greater than the given threshold. Figure~\ref{fig:candidatesoutputlayout} shows the mapping of probing sets to their candidate pairs and the output result array (e.g. only ($r_{3}$, $s_{4}$) pair is similar). Every even index position in $C_{O}$ corresponds to a probe set id and every odd one to a candidates offset. In the figure, since $C_{O}[0]=r_{1}$ and $C_{O}[1]=1$, $r_{1}$ should be compared against only the first entry of $C$. Since  $C_{O}[3]-C_{O}[1]=0$ and $C_{O}[2]=r_{2}$, $r_{2}$ does not participate in any candidate pair. Finally, since  $C_{O}[5]-C_{3}[1]=2$ and $C_{O}[2]=r_{3}$, $r_{3}$ should be verified against the next two sets in $C$.
	
	\textbf{Memory restrictions.} In our data layout, most memory space is required to store the array of candidate set IDs $C$, which is of quadratic space complexity. Depending on the dataset and threshold value, the overall number of candidate pairs could reach billions. Considering that a set id is a 4-byte integer, $||C||$ space could be of several gigabytes. In addition, $||O||=\frac{||C||}{4}$ space is also required to store the output result (assuming that 0 and 1 take a byte rather than a bit). Due to the limited device memory, the host iteratively transfers as many candidates as the device memory can handle (so that both candidate set IDs and output fit into the  global memory). This favors the overlap between both ends, as the host builds the next chunk of candidate pairs and the device conducts the verification in parallel.

	\subsubsection{Verification alternatives}
	We define three main kernel-level concurrency layers or dimensions of our problem as follows: \textit{grid layout} (\textbf{GL}), which corresponds to thread execution; \textit{memory hierarchy} (\textbf{MH}), which corresponds to efficient exploitation of the fastest on-chip memories; and last \textit{output writing} (\textbf{OW}), which deals with result output. The latter is also distinguished into two cases depending on the type of querying being performed: \textit{output count} (\textbf{OC}) for an aggregate query and \textit{output select} (\textbf{OS}) for a full select of the similar pairs query.
	
	\begin{figure}[tb!]
		\centering
		\includegraphics[width=0.9\linewidth]{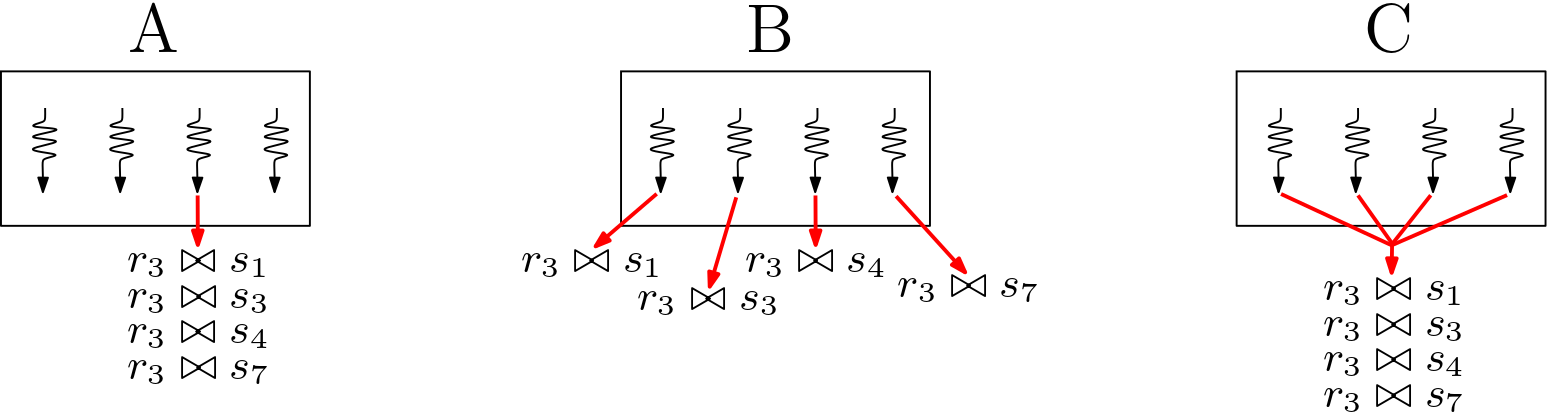}
		\caption{Thread workload per alternative}
		\label{fig:alternatives}
	\end{figure}

	These layers are tightly coupled and often intertwined, which means that certain options on a layer can rule out available options on the next ones. To investigate the impact of each available layer option, we present three alternative scenarios below, which differ in the workload assigned to a single thread, as shown in Figure~\ref{fig:alternatives}.

	\textbf{Alternative A.} In our first alternative, the workload we  assign to each thread is a  probing set and the evaluation of all its corresponding candidate pairs, i.e., a single thread becomes responsible for the verification of all candidate pairs involving a specific probing set.
	
	\noindent\underline{GL}: We launch a 1D grid of 1D blocks, with the overall number of threads executed across all blocks ($T$) being equal to the input set collection size ($R$). Each thread is responsible for a probing set $r_{i}$ and conducts all the joins with the corresponding candidates $s_{j}$.
	
	\noindent\underline{MH}: In this alternative, we do not use the shared memory. An option could be each thread to load the corresponding probing set $r_{i}$ to the shared memory, and then to access every candidate set $s_{j}$ from global memory. Thus, a thread does not access global memory for $r_{i}$ during the verification of each candidate pair. Since there is no fixed set size, blocks which handle sets of thousand tokens require larger amount of shared memory. For example, a thread block of 32 threads and average probing set size equal to 1000, would require $32\times1000\times4=128$KB of shared memory which exceeds the maximum of 48KB that modern GPUs can support. In such cases, an adaptive approach must be followed where the thread block size is limited to allow for proper execution. In summary, the option to load $r_i$ in shared memory implicitly defines the block size and gives rise to further challenges; thus is avoided in this work.
	
	\noindent\underline{OC}: As every thread verifies  its own candidate pairs independently, it can also count the amount of pairs satisfying the threshold using a register. After finishing the verification, each counter can be stored in shared memory in order for a fast reduction on block level to be performed. The result of each block is then  stored in global memory for a grid level reduction to output the global count. The amount of shared memory required depends on the block size, but it is small (e.g., for 32 threads per block we need 128 bytes).
	
	\noindent\underline{OS}: Having allocated the memory required for output array $O$, a device thread updates specific cells of the array. Incorporating the shared memory in this output is not straightforward because each thread does not know beforehand the length of its output pairs. Hence, we have to allocate shared memory for the worst case scenario per thread. However this is practically impossible with the current GL. For example, if a block has 100000 candidate pairs to verify, we have to allocate 100KB of shared memory, which exceeds the maximum allowed space. As a result, the shared memory cannot be employed to speed-up the output generation.
	
	%solution: divide equally shared memory to hold output per thread, e.g. 64KB shared memory , 32 threads per block leads to 2KB per thread or $\approx$500 output pairs upperbound.
	
	%Memory latency is hidden by very fast context switching; when a memory fetch is issued while processing one subset of data elements,

	\textbf{Alternative B.} In this technique, we allocate less work to each thread by shifting the workload of a single probing set from a single thread to a single thread block. By assigning the comparisons referring to a probing set to a thread block,  threads evaluate only a portion of the candidate pairs in parallel. The main benefit of this alternative is that the workload of threads within a block is more evenly distributed.
	
	\noindent\underline{GL}: We launch a 1D grid of 1D blocks, with the number of blocks being equal to the input set collection size. Each thread block is responsible for a probing set and each thread is assigned with a portion of candidate pairs to verify.
	
	\noindent\underline{MH}: First, the block threads load the corresponding probing set $r_{i}$ to  shared memory, then each thread verifies a portion of candidate pairs by accessing the corresponding candidate sets $s_{j}$ from global memory. Because we use shared memory for one probing set per block, unlike alternative A, the maximum supported probing set size also increases.
	
	\noindent\underline{OC/OS}: Same as Alternative A.

	\textbf{Alternative C.} In our previous scenario, we try to improve performance on the warp level. We further extend the rationale of alternative B, and more specifically, each block is assigned with a probing set but with the difference that all the block threads cooperate to evaluate a candidate pair using the intersect path algorithm proposed in \cite{green2014fast}. This further mitigates the problem of balancing, since the threads do not only become responsible for an equal number of candidates, but also perform a roughly equal number of operations.
	
	%The  first  partitioning  splits  the  two  input  arrays
	%into  equal-sized  units  that  can  be  merged  independently  by
	%the multiple GPU multiprocessors.
	%
	% Essentially, input lists are divided into  the workload is evenly divided on each thread

	\noindent\underline{GL}: We launch a 1D grid of 1D blocks, with a number of thread blocks equal to the input set collection size. Each thread block is responsible for a probing set and multiple threads contribute to each candidate pair verification. A control thread outputs the result to global memory.
	
	\noindent\underline{MH}: Extending alternative B, due to thread cooperation, by default we also load candidate sets to shared memory. If there is not enough space to  hold all candidate sets, we load data in chunks, perform the verification, and then proceed to the next chunk.
	
	\noindent\underline{OC}: Since all threads within a block cooperate to verify a candidate pair, we assign only a single thread to increment the block's counter. Thus there is no need for a thread block counter reduction. However, grid level reduction is still required.
	
	\noindent\underline{OS}: The same applies for updating the output array $O$. If a candidate pair meets the threshold, only a single thread updates the corresponding array cell.

	\section{Implementation Issues}
	\label{sec:impl}
	
	%In this section, we provide the main technical details that significantly affect the performance and the design approach of which is not trivial.
	
	\subsection{Host Details}
	
	Our framework leverages the work of Mann \cite{mann:ssjoin}. The main difference is that
	we migrate the verification phase from the host to the device side via a multi-threaded implementation, which gives rise to the issues discussed in this section.
	
	\subsubsection{Candidate Serialization}
	
	% gives prominence
	The need to transfer candidate pairs to device memory highlights the necessity of efficient serialization methods. Our goal is for the device to avoid complex global memory accesses. Therefore, the  host is responsible for storing the candidates of a probe set in successive memory addresses. We list our options for serializing candidates $C$ as follows:
	\begin{enumerate}
		\item Use a sequence container such as \textit{std::vector} and push back every new candidate. The main drawback is the extra memory checks on insertion to determine if a reallocation is required.
		\item Prior reserve memory space for \textit{std::vector} to avoid memory checks.
		\item Use primitive arrays and handle memory operations manually.	
		\item Use a map structure where a key is an integer, i.e. the probe set ID, and its value is a \textit{std::vector} containing the corresponding candidates IDs.
	\end{enumerate}
	
	As a complement to $C$, a separate array $C_{O}$ to delimit candidate pairs is required. Moreover, the tokens that we insert to $C_{O}$ are pairs consisting of a probe set ID and its corresponding offset on $C$. Omitting the probe set ID and inserting the candidate offset by itself, thus reducing the $C_{O}$ size, implies that the probe set ID should be capable to be \textit{extracted} from the index of $C_{O}$. For that to be possible, continuous probe sets IDs in ascending order must be processed, which might not always be the case.
	
	The use of map is an intermediate stage to group together in memory probe set candidates. Before invoking the device, we must iterate the map and serialize every candidates list to construct the final $C$, and also update $C_{O}$ per iteration.
	
	As will be shown in our experiments, primitive arrays, i.e., the third option listed above, perform better than \textit{std::vector}, i.e., the first two options, and are adequate for ALL and PPJoin that produce the full candidate set for a single probing set in a single phase. For GRP, which employs two phases during candidate generation,
	a map structure is necessary if the full verification phase is delegated to the GPU.

	\subsubsection{Thread \& Memory Management}
	As stated in Section~\ref{sec:approach}, our framework consists of three threads. The indexing/filtering thread $H_{0}$ reserves memory space beforehand and serializes candidates. When the device maximum memory for candidate pairs $M_{c}$ is filled, $H_{0}$ triggers $H_{1}$, the device handler thread, and assigns  a pointer to the current candidates to it. In parallel, $H_{1}$ allocates the required device memory space, copies the candidates array to device and launches the join kernel code. Meanwhile, $H_{0}$ continues to build the next chunk of candidates. As soon as the join kernel finishes, if an aggregation on top of the join is requested, $H_{1}$ launches immediately a separate kernel to perform a count reduction. In case of output the actual pairs, $H_{1}$ starts the $H_{2}$, which post-process $O$. When $H_{2}$ finishes, the memory reserved for the respective candidates is freed. The same steps are repeated until no candidate pairs are left to be verified.
	
	%mention that $M_{h}>=2M_{d}$

	\subsubsection{GroupJoin Work Split}
	
	The three best-performing algorithms examined can be divided into two categories: those which generate every candidate pair in one phase, i.e. ALL and PPJ, and the one, GRP, which requires an extra phase (group expanding) to output all candidate pairs.
	
	For each probe set in ALL and PPJ, candidates are guaranteed to be stored in successive memory addresses. Thus, we serialize candidates using primitive arrays. In contrast, GRP generates candidates that are intertwined due to the expanding phase. Therefore storing candidates in a map structure is required. However, in our testings serializing the map adds extra overhead, rendering this option unfeasible.
	
	We choose to split the work for GRP, and for this technique, allocate part of the verification to the CPU as well. We assign the verification of every candidate pair generated in the first phase to the device. Hence, we serialize candidates from this phase in primitive arrays and transfer them to the device. Every candidate pair that emerges from the second phase, i.e. group expanding, is left to be verified in the host side by $H_{0}$.
	
	By splitting the work, we alleviate overall performance as shown in Section~\ref{sec:eval}. In spite of this gain, transferring the whole GRP verification workload to device remains a challenge and would further improve the performance.
	
	\subsection{Device Details}
	
	%Achieving good performance on the device requires fine-code tuning and also the right parameter setting. Furthermore, we analyze our block launch parameters, present our intersect path algorithm adaptation for supporting the third verification alternative, and describe a fast reduction method to minimize the global memory footprint.
	
	\subsubsection{Block size}
	We launch grids composed of 1D blocks. The block size $B$ must be a power of 2 for reduction on the shared memory to work properly. We prefer $B=32$ since a \textit{warp} can be considered as the CPU thread equivalent. However in our evaluation with different block sizes, we show a correlation between block size and set size, and increasing the block size should be considered when the third verification alternative is the best performing one.

	\subsubsection{Merge and Intersect Path}
	
	Computing a list intersection can be derived from a list merge operation. An efficient parallel merging algorithm for GPUs is Merge Path\cite{green2012gpu}. Given two sorted lists $A$ and $B$, Merge Path considers the order in which elements are merged, which is equivalent to the traversal of a grid, noted as \textit{Merge Matrix}, of size $|A|\times|B|$. Beginning from the top left corner of the grid, the path can only move to the right if $A[i] \geq B[j]$ or downwards otherwise, until it eventually reaches the bottom right corner.
	
	There are two partitioning stages, one on kernel grid level and the other on block level. On grid level, equidistant cross diagonals are placed on the Merge Matrix. Using binary search, the point of intersection for a cross diagonal and the path is found. As a result, each SM is assigned to merge non-overlapping portions of the input. On block level, threads cooperate in loading the required list portions on shared memory and then merge them in global memory.% binary search doesn't compute the entire path but instead only partitioning points.
	
	By modifying Merge Path in~\cite{green2014fast}, the authors propose a fast list intersection algorithm, called Intersect Path. They introduce a new diagonal path move, if $A[i] = B[j]$. The same partitioning stages still hold. Each SM outputs a portion of the intersection to global memory.
	
	Because in our approach thread blocks verify whole candidate pairs, we have to modify Intersect Path accordingly. Thus we perform both partitioning stages on block level.
	
	\begin{figure}[tb!]
		\centering
		\includegraphics[width=0.57	\linewidth]{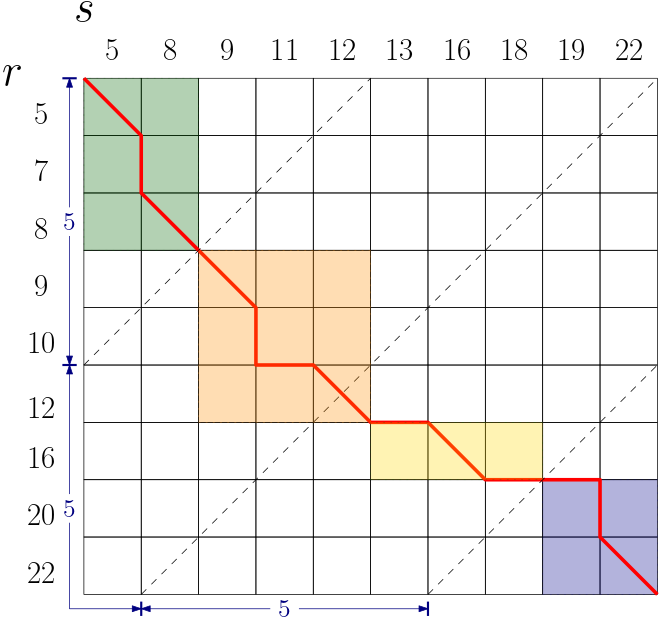}
		\caption{Intersect path example.}
		\label{fig:intersectpath}
	\end{figure}

	\subsubsection{Set Intersection Count}
	In order to verify a candidate pair, threads must calculate the intersection of two sets. Each thread in alternatives A and B,  independently performs a merge-like loop and counts the number of intersects. On the other hand, in alternative C, threads collaborate to output the intersection. Since our problem can be reduced to finding the intersection count of two sets, we use a modified Intersect Path algorithm to divide the workload between block threads, as mentioned above.
	
	Cross diagonals are equally placed apart at $\lceil\frac{|r_{i}| + |s_{j}|}{B} \rceil$ \cite{green2014fast}. Each thread is assigned with a partition with starting point the intersection of the path and the corresponding diagonal. In Figure~\ref{fig:intersectpath}, we show an example of Intersect Path using $B=4$ threads (each color corresponds to a different thread). Starting from the top left corner, cross diagonals are placed apart 5-hops away, where hops are along the axes. Each thread calculates independently its partition intersection count and stores it in registers memory. When finished, intersection counts are copied in shared memory for a fast reduction to output the overall intersection.
	%mention exception when cross diagonal intersects the path inside the square

	\subsubsection{Count Reduction}
	
	Our primary focus is to minimize the global memory transactions. Whenever possible, we use the registers to store counts per thread. Similarly, we store every thread count to shared  memory and use warp shuffle functions for fast reduction. Hence, when performing an aggregation on top of the join, only a single write to global memory is required per block. To output the final count, we use \textit{thrust::reduce} on the global memory-resident intermediate counts.
	
	\section{Evaluation}
	
	%In this section, we present and analyze our experimental results. To further support our findings, we profile the device behavior on several \textit{representative} experiments. %interesting experiments?
The goals of our experiments are threefold:	(i) to show the extent to which the verification phase delegated to GPU is hidden (overlapped) by the index and filtering phases running on the CPU; (ii) to give concrete evidence about the speed-ups achieved in practice, and (iii) to provide explanations about the observed behavior.

	\label{sec:eval}
	
	%transfer times negligible
	%impact of candidate serialization
	%mention cpu iteration- serialization tradeoff,

	%\textbf{Results.} candidate time = prefix index (built incrementally). verification time includes scanning the candidate set and verifying each candidate. join time is sum of candidate and verification and does not include preprocessing time.  true positives must be verified
	
	%	\item formulas,  (?) $\rightarrow$ $r_{i} \bowtie s_{j}, (i, j \in N) | s_{j} \subseteq S, j \in N| sim(r_{i}, s_{j}) \geq t$
	
	\subsection{Experiment setting}
	
	The experiments were conducted on a machine with an Intel i7 5820k clocked at 3.3GHz, 32 GB RAM at 2400MHz and an NVIDIA Titan XP. This GPU has 3840 CUDA cores, 12 GB of global memory and a 384-bit memory bus width.
	
	The overall runtime, noted as \textit{join} time, is the composition of candidate generation and serialization performed by $H_{0}$ host thread, and the verification conducted by the device.  We refer to the former as \textit{index/filtering} time and to the latter as \textit{verification} time. We do not include any data preprocessing time spent for tokenization and de-duplication, which are perfomed exactly as in \cite{mann:ssjoin}.	
	We conduct experiments for all datasets using ten thresholds in the range [0.5, 0.95]. We focus on self-joins using the Jaccard similarity and perform an aggregation on top of the set similarity join.
	The reported time for each experiment is an average over 3 independent runs (no significant deviation was observed). We measure the \textit{index/filtering} and total \textit{join} time with the \textit{std::chrono} library. For the \textit{verification} time we use the CUDA event API. The times for allocating device memory and transferring chunks of candidates to the device were negligible for all the experiments and furthermore were completely hidden due to the execution overlap.

%\subsubsection{Datasets}
	
\begin{figure}[tb!]
	\centering
	\includegraphics[width=0.8	\linewidth]{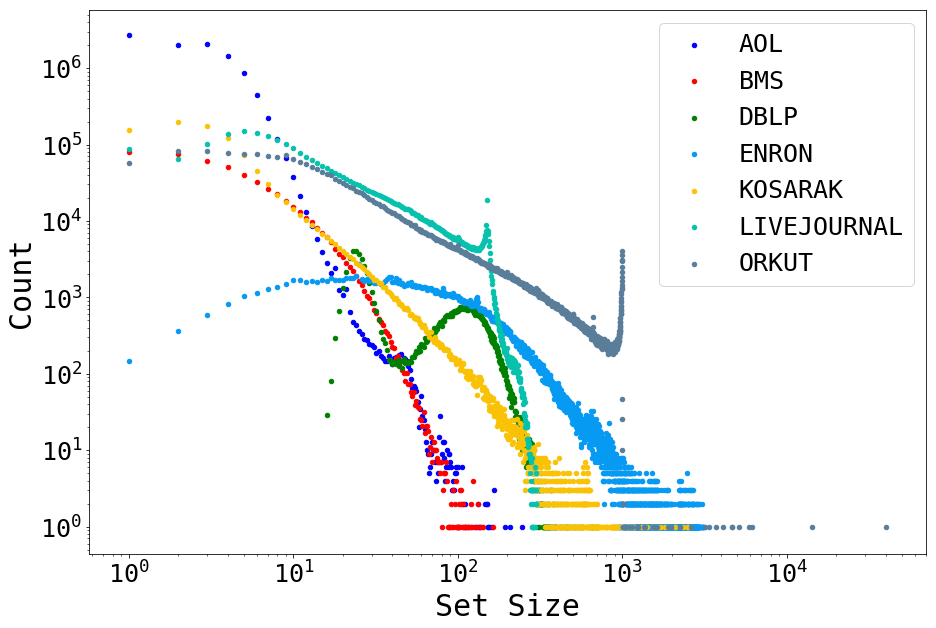}
	\caption{Datasets set size distribution}
	\label{fig:datasets_distro}
	\end{figure}

	We experiment with seven real world datasets that were also  employed in \cite{mann:ssjoin}.  Table~\ref{tab:datasets} shows an overview of each dataset characteristics. Some datasets follow a Zipf-like distribution of set sizes, as shown in Figure~\ref{fig:datasets_distro}, but in general, the distribution types differ. A summary of each dataset (adapted from~\cite{mann:ssjoin}) is  as follows:
%Hence, a  single tokenization technique is applied per dataset. Duplicate tokens are deduplicated using a counter that is appended to each duplicate, thus increasing the number of overall tokens.
	
	\begin{table}[tb!]
		\renewcommand{\arraystretch}{1.3}
		\caption{Datasets characteristics (*DBLP is further increased later)}
		\label{tab:datasets}
		\centering
		\begin{tabular}{c||c||c||c}
			Dataset 	& Cardinality 	& Average set size	&\# diff tokens \\
			\hline
			AOL			& $1.0 \cdot 10^7$  	& 3 		& $3.9 \cdot 10^6$ \\
			BMS-POS		& $3.2 \cdot 10^5$ 		& 6.5		& 1657 \\
			DBLP*		& $1.0 \cdot 10^5$ 		& 88		& 7205 \\
			ENRON		& $2.5 \cdot 10^5$		& 135		& $1.1 \cdot 10^6$ \\
			KOSARAK		& $6.1 \cdot 10^5$		& 8			& $4.1 \cdot 10^4$ \\
			LIVEJOURNAL	& $3.1 \cdot 10^6$      & 36.5		& $7.5 \cdot 10^6$ \\
			ORKUT		& $2.7 \cdot 10^6$      & 120		& $8.7 \cdot 10^6$ \\
		\end{tabular}
	\end{table}
	\noindent
	
	\begin{figure*}[bt!]
		\centering
		\subfigure[AOL]
		{
			\includegraphics[width=0.23\linewidth]{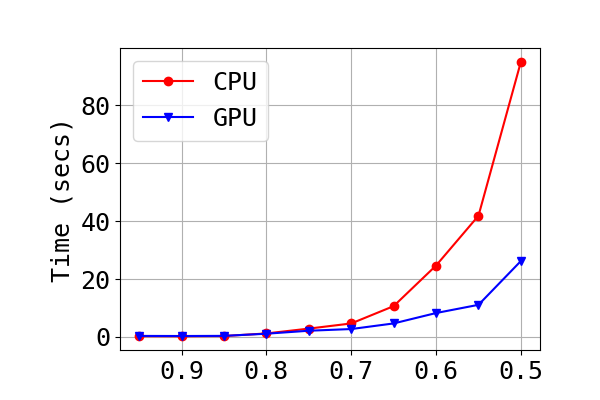}
			\label{fig:aol_comparison_ver}
		}
		\subfigure[BMS]
		{
			\includegraphics[width=0.23\linewidth]{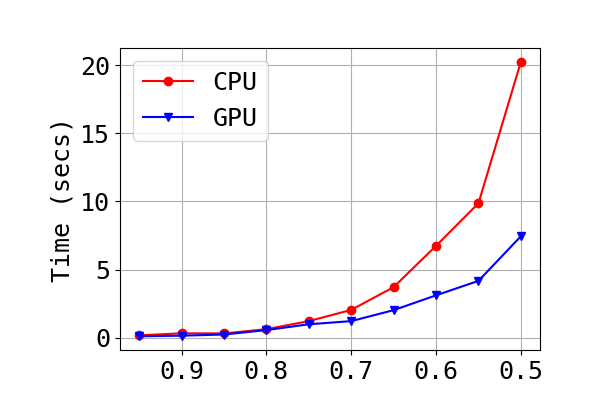}
			\label{fig:bms_comparison_ver}
		}
		\subfigure[DBLP]
		{
			\includegraphics[width=0.23\linewidth]{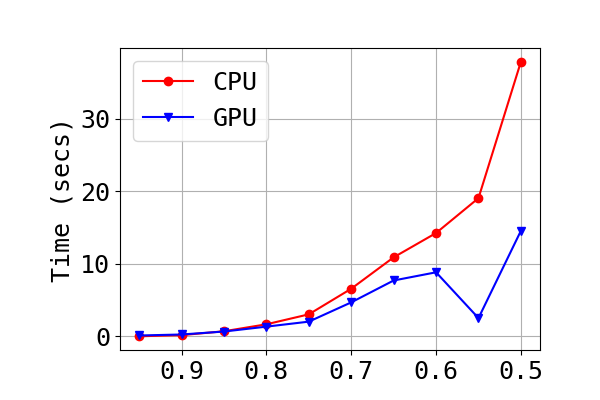}
			\label{fig:dblp_comparison_ver}
		}
		\subfigure[ENRON]
		{
			\includegraphics[width=0.23\linewidth]{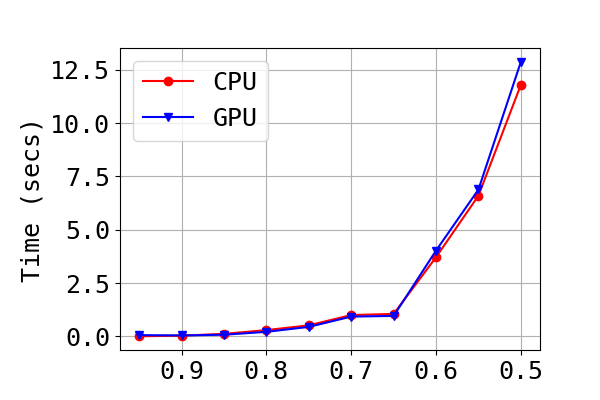}
			\label{fig:enron_comparison_ver}
		}
		\subfigure[KOSARAK]
		{
					\includegraphics[width=0.23\linewidth]{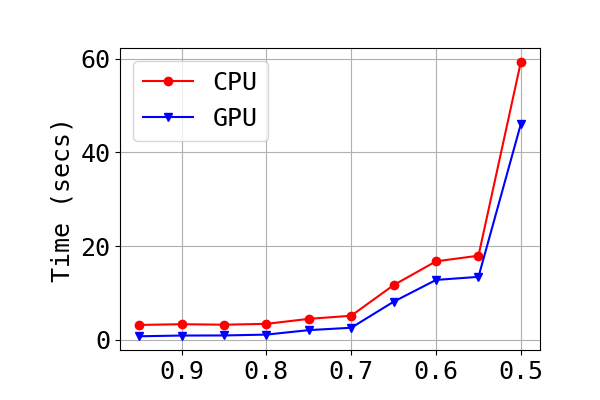}
			\label{fig:kosarak_comparison_ver}
		}
		\subfigure[LIVEJOURNAL]
		{
			\includegraphics[width=0.23\linewidth]{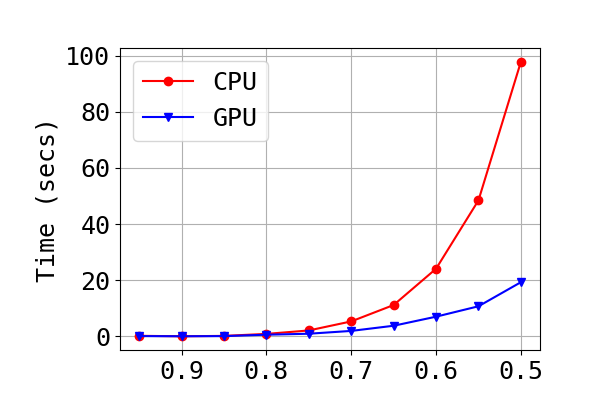}
			\label{fig:livejournal_comparison_ver}
		}
		\subfigure[ORKUT]
		{
			\includegraphics[width=0.23\linewidth]{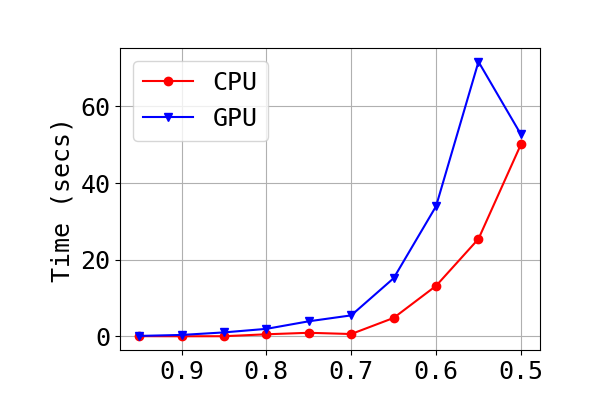}
			\label{fig:orkut_comparison_ver}
		}
	\caption{Comparison between the best verification times of the CPU and an unoptimized GPU execution ($B=32$ and $M_{c}=4GB$) for different thresholds.}
		\label{fig:cpu_gpu_comparison_ver}
	\end{figure*}

	\begin{figure*}[bt!]
		\centering
		\subfigure[AOL]
		{
			\includegraphics[width=0.23\linewidth]{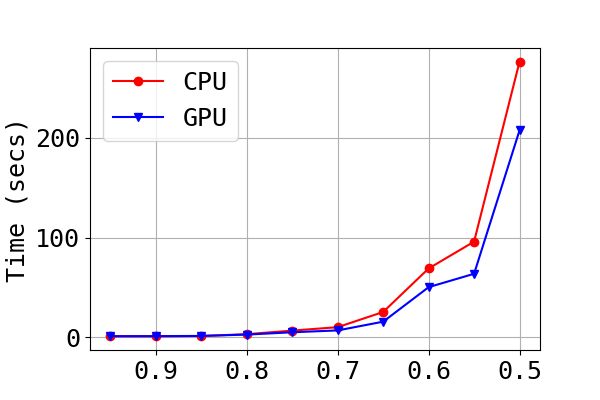}
			\label{fig:aol_comparison}
		}
		\subfigure[BMS]
		{
			\includegraphics[width=0.23\linewidth]{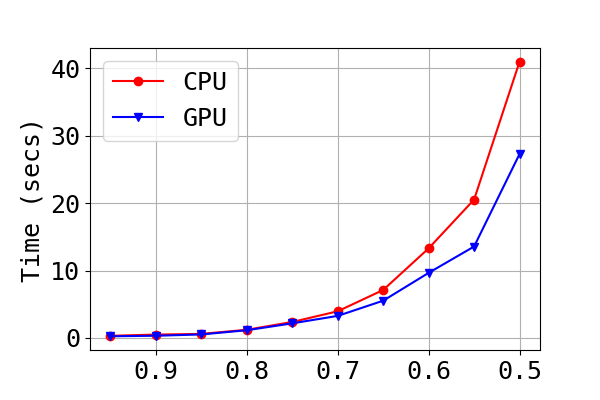}
			\label{fig:bms_comparison}
		}
		\subfigure[DBLP]
		{
			\includegraphics[width=0.23\linewidth]{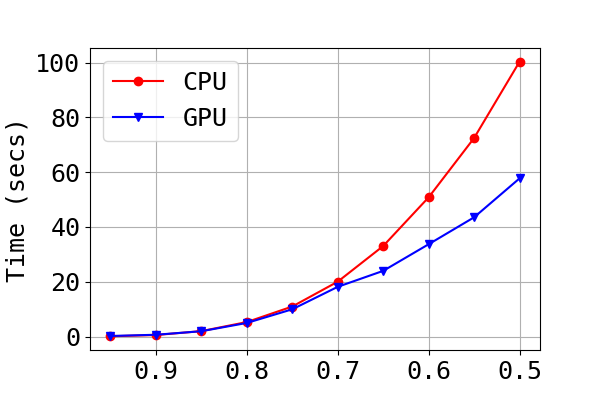}
			\label{fig:dblp_comparison}
		}
		\subfigure[ENRON]
		{
			\includegraphics[width=0.23\linewidth]{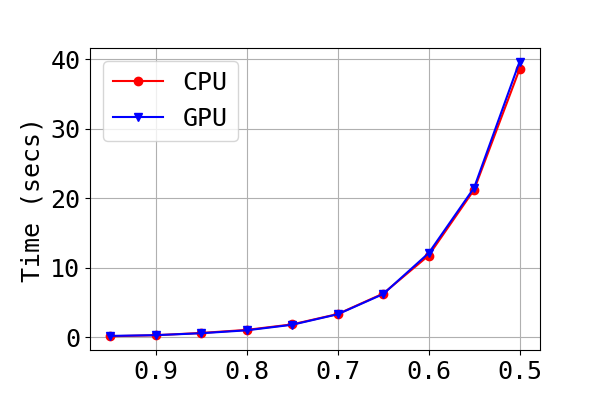}
			\label{fig:enron_comparison}
		}
		\subfigure[KOSARAK]
		{
			\includegraphics[width=0.23\linewidth]{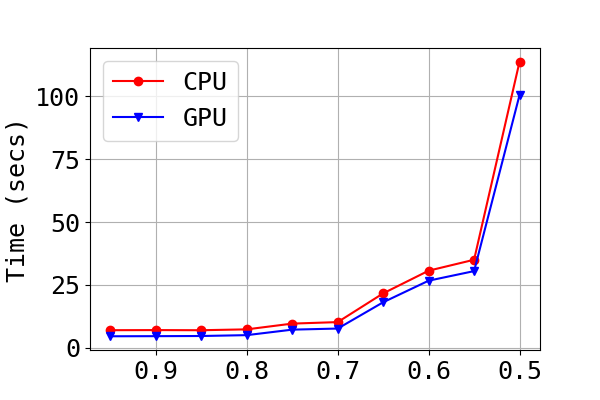}
			\label{fig:kosarak_comparison}
		}
		\subfigure[LIVEJOURNAL]
		{
			\includegraphics[width=0.23\linewidth]{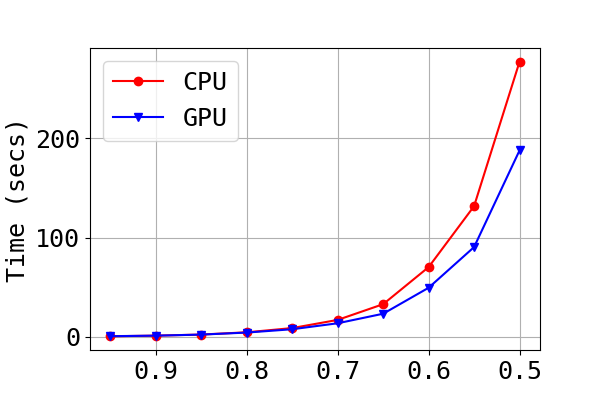}
			\label{fig:livejournal_comparison}
		}
		\subfigure[ORKUT]
		{
			\includegraphics[width=0.23\linewidth]{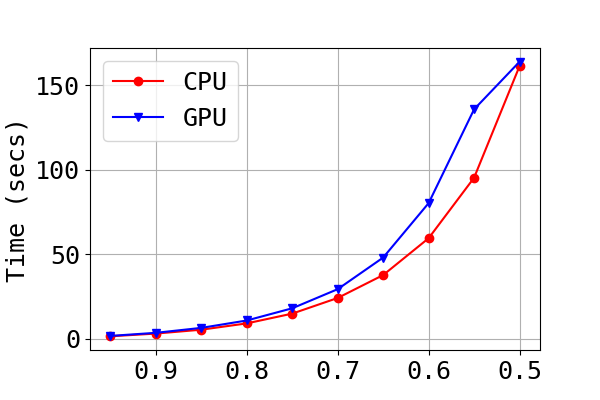}
			\label{fig:orkut_comparison}
		}
		\caption{Comparison between the best join times of the CPU and an unoptimized GPU execution ($B=32$ and $M_{c}=4GB$) for different thresholds.}
		\label{fig:cpu_gpu_comparison}
	\end{figure*}

	\begin{figure*}[tb!]
		\centering
		\subfigure[200k DBLP]
		{
			\includegraphics[width=0.23\linewidth]{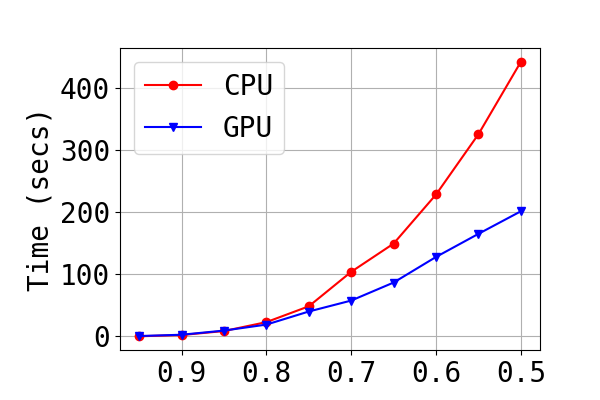}
			\label{fig:scalability_dblp200k}
		}
		\subfigure[300k DBLP]
		{
			\includegraphics[width=0.23\linewidth]{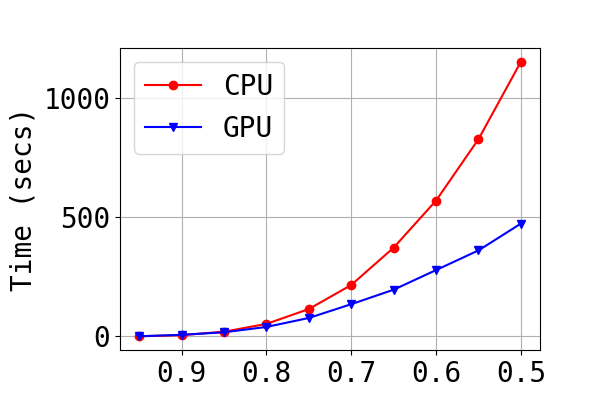}
			\label{fig:scalability_dblp300k}
		}
		\subfigure[1M DBLP]
		{
			\includegraphics[width=0.23\linewidth]{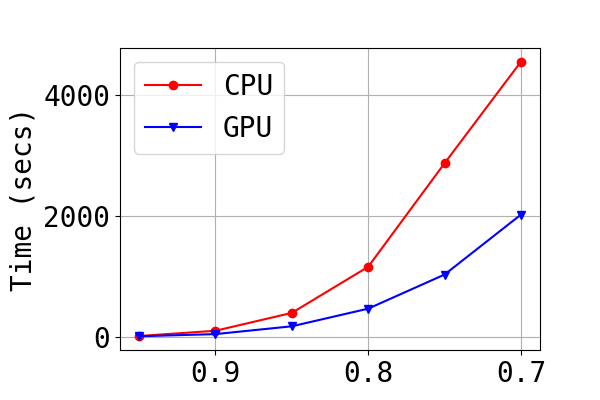}
			\label{fig:scalability_dblp1M}
		}
		\subfigure[Complete DBLP (6.1M)]
		{
			\includegraphics[width=0.23\linewidth]{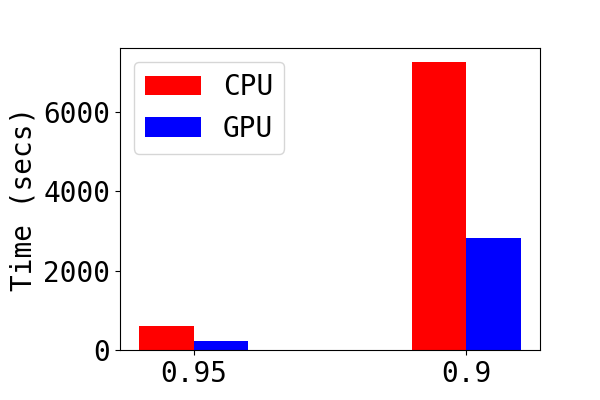}
			\label{fig:scalability_dblp_complete}
		}
		\caption{Comparison between the best times of the CPU and an unoptimized GPU execution for larger datasets  for different thresholds.}
		\label{fig:scalability}
	\end{figure*}

	\textbf{AOL:} query log data from the AOL search engine. Each set represents a query string and its tokens are search terms.
	
	\textbf{BMS-POL:} purchase data from an e-shop. Each set represents a purchase and its tokens are product categories in that purchase.
	
	\textbf{DBLP:} article data from DBLP bibliography. Each set represents a publication and its tokens are character $2$-grams of the respective concatenated title and author strings.
	
	\textbf{ENRON:} real e-mail data. Each set represents an e-mail and its tokens are words from either the subject or the body field.
	
	\textbf{KOSARAK:} click-stream data from a Hungarian on-line news portal. Each set represents a user behavior and its tokens are links clicked by that user.
	
	\textbf{LIVEJOURNAL:} social media data from LiveJournal. Each set represents a user and its tokens are interests of that user.
	
	\textbf{ORKUT:} social media data form ORKUT network. Each set represents a user and its tokens are group memberships of that user.

	\subsection{Main Experiments}

	We compare our hybrid framework to the CPU standalone implementation of Mann\cite{mann:ssjoin}. In Figure \ref{fig:cpu_gpu_comparison_ver}, we show the speed-ups that can be achieved during the verification phase.  These speed-ups can be up to more than 5X in our experiments. Nevertheless, in practice, it is more important to investigate the total response time, which inludes both the filtering/indexing and the verification phase.
	
	In Figure~\ref{fig:cpu_gpu_comparison}, we present the best join times measured for both. Each time reported for the CPU is the overall best of the three algorithms execution and therefore the best we can achieve in our setup. Respectively for the GPU, each time reported is the best among unoptimized executions ($B=32$, $M_{c}=4$ GB) of the three algorithms and the three alternatives. Thus, performance can be further improved, especially on datasets such as ENRON (Figure~\ref{fig:enron_comparison}) and ORKUT (Figure~\ref{fig:orkut_comparison}) where the GPU performs similar or worse than the CPU.
	
	As shown in Figure~\ref{fig:cpu_gpu_comparison}, for every dataset on large thresholds, i.e. the threshold range is in [0.7, 0.95], the GPU does not yield any performance speedup. Given the fact that in large thresholds the number of candidate pairs, hence the memory required to store them, is quite smaller than the device memory $M_{c}$, the GPU remains idle during the candidate generation phase, only to be invoked once before the process finishes. On the other hand, billions of candidate pairs are generated when using smaller thresholds ([0.5, 0.65]). As a result, the GPU is invoked several times leading to an execution overlap and therefore to faster join times.
	
To further support this conclusion, we run our techniques over larger DBLP datasets, as shown in 	Figure~\ref{fig:scalability} (where we present settings in which CPU takes up to an hour approximately). We can see that the speed-ups are much more evident, e.g., 2.6X in Figure \ref{fig:scalability_dblp_complete}, and are tangible even for larger thresholds, where the candidate pairs are tens of billions. In Figure \ref{fig:scalability_dblp1M}, the candidates are 3.5B, 17B and 77B for thresholds 0.9, 0.8, and 0.7, respectively, and there are clear benefits for the last two settings.

	\begin{table}[tb!]
	\renewcommand{\arraystretch}{1.1}
		\caption{GPU join time decomposition for processing the complete DBLP dataset (in secs)}
		\label{tab:overlap}
		\centering	
		\begin{tabular}{c||c||c c | c ||c}
$t$ & join  & \multicolumn{2}{c|}{index/filtering} & verification & $||C ||$\\
 &   & filtering & serialization & & \\ \hline
0.95 & 233 & 134 & 96 & 92 & 72.7GB\\
0.9 & 2815 & 1892 & 921 & 698 & 0.56TB\\
0.85 & 11367 & 8935 & 2430 & 2311 & 1.8TB\\
		\end{tabular}
	\end{table}
	\noindent

Further, we drill-down on the GPU join time, as shown in Table \ref{tab:overlap}, where it is shown that the GPU join time is solely attributed to the index/filtering time; the join time is roughly equal to the sum of filtering, index building and serialization. This means that our GPGPU scheme that assigns verification only to the GPU has reached its maximum potential.

\textbf{Key result:} Our GPGPU solution manages to hide the impact of verification phase on the running time of the similarity join, when (i) the candidates are in the order of tens of billions at least and (ii) the index/filtering and verification phases are intertwined.

	\subsection{Algorithm Performance}
			
	\begin{table}[tb!]
		\caption{Number of datasets in which each algorithm is the best per threshold}
		\label{tab:algs}
		\centering	
		\begin{tabular}{c|p{0.04\columnwidth}p{0.04\columnwidth}p{0.04\columnwidth}p{0.04\columnwidth}p{0.04\columnwidth}p{0.04\columnwidth}p{0.04\columnwidth}p{0.04\columnwidth}p{0.04\columnwidth}p{0.04\columnwidth}}
				& 0.95 & 0.90 & 0.85 & 0.80 & 0.75 & 0.7 & 0.65 & 0.6 & 0.55 & 0.5 \\ \hline
ALL & 1 & 1 & 0 & 1 & 2 & 3 & 4 & 4 & 5 & 3 \\
PPJ & 2 & 2 & 2 & 4 & 4 & 3 & 2 & 2 & 1 & 3 \\
GRP & 4 & 4 & 5 & 2& 1 & 1 & 1 & 1 & 1 & 1 \\
		\end{tabular}
	\end{table}
	\noindent

Table \ref{tab:algs} shows which algorithm
exhibited the best performance in Figure~\ref{fig:cpu_gpu_comparison} on the GPU. We can observe, that as in the CPU comparison in \cite{mann:ssjoin}, there is no algorithm that dominates the others. However, ALL favors low thresholds, PPJ mid-range and GRP the high ones, especially for the datasets with small average size.

Next, we discuss algorithm behavior issues in relation to the three datasets, where GPU does not show tangible benefits in Figure~\ref{fig:cpu_gpu_comparison} even for relatively low thresholds, namely ENRON, KOSARAK and ORKUT.

	\subsubsection{Execution overlap and the role of $M_c$ }
	
	The most significant gain due to the device stems from the execution overlap between indexing and verification. ALL invokes the device earlier and more frequently because of its fast candidate generation on all datasets. PPJ and GRP have higher index times, generate less candidates and as a result invoke the device less frequently.	

\begin{figure}[tb!]

		\subfigure[ENRON]
		{
			\includegraphics[width=.45\columnwidth]{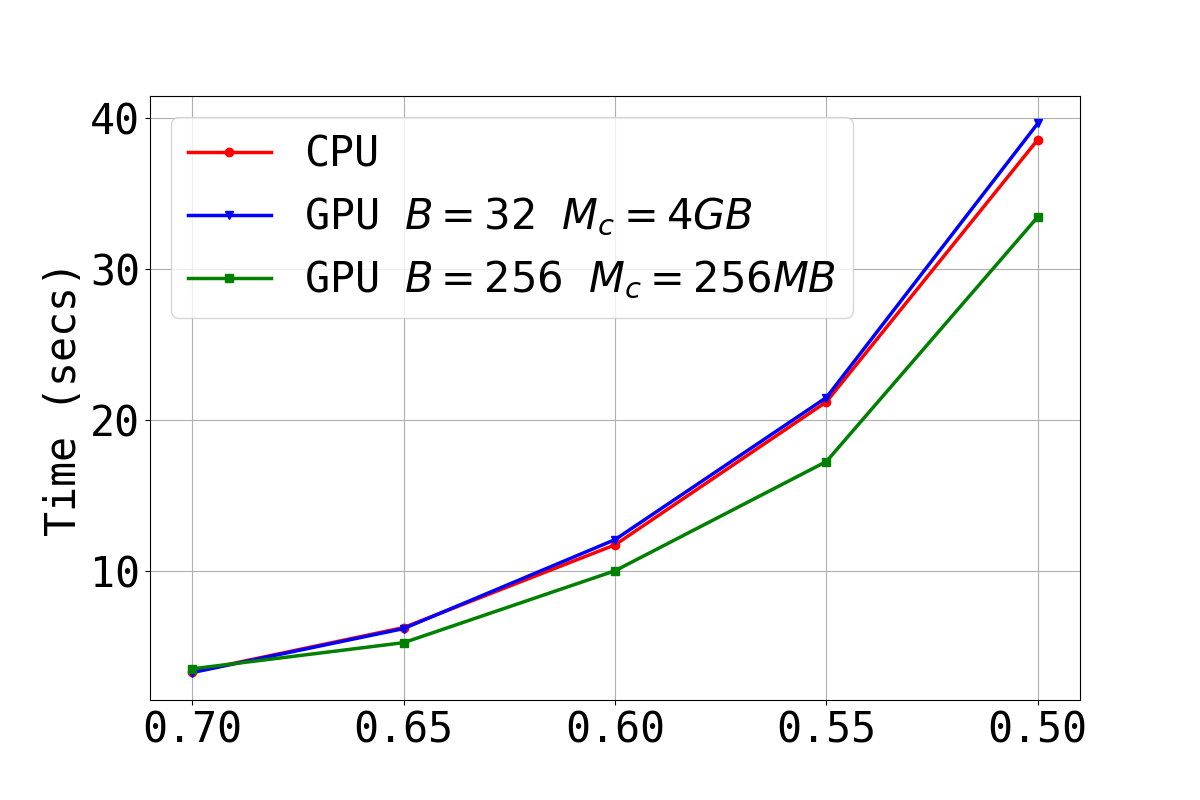}
			\label{fig:enron_execution_overlap}
		}
		\hfill
		\subfigure[ORKUT]
		{
			\includegraphics[width=0.45\columnwidth]{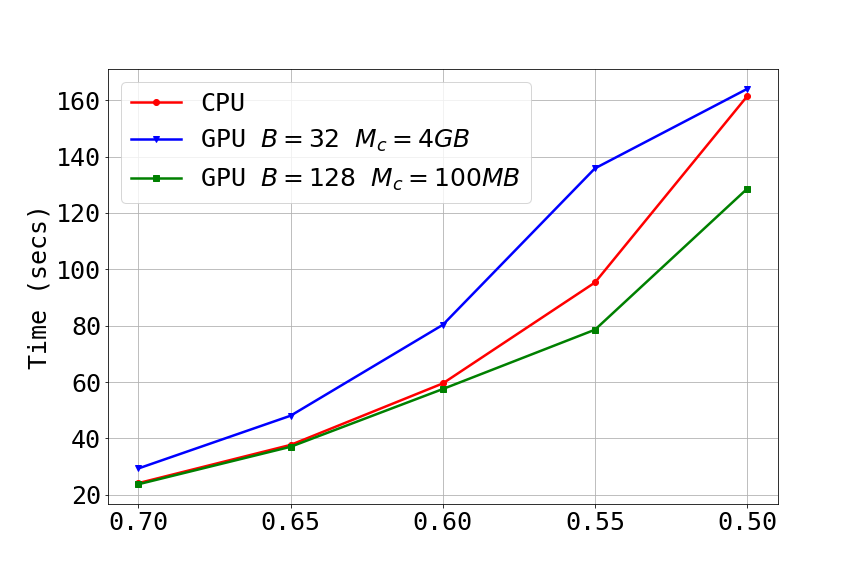}
			\label{fig:orkut_execution_overlap}
		}
		\caption{Impact of execution overlap on the GPU performance.}
		\label{fig:execution_overlap}
	\end{figure}

	To exploit the execution overlap, and therefore keep the device busy most of the time, defining the appropriate $M_{c}$ is necessary. We examine the ENRON and ORKUT datasets. For these two datasets, all three algorithms have similar index times. The overall best algorithm for the  standalone CPU execution is PPJ since it generates less candidates than ALL in the same indexing time frame and does not include any candidate expanding during the verification phase as GRP does. Figure~\ref{fig:enron_execution_overlap} and Figure~\ref{fig:orkut_execution_overlap} show the join time for the ENRON and ORKUT dataset respectively, using PPJ as the index algorithm. By decreasing $M_{c}$, and by fine-tuning $B$ as will be shown in the device performance section, we increase overlapped execution between CPU and GPU and manage to hide the verification time in the execution overlap.
	
	\subsubsection{GroupJoin Issues}
	
\begin{figure}[tb!]
		\subfigure[KOSARAK, alternative B]
		{
			\includegraphics[width=0.45\columnwidth]{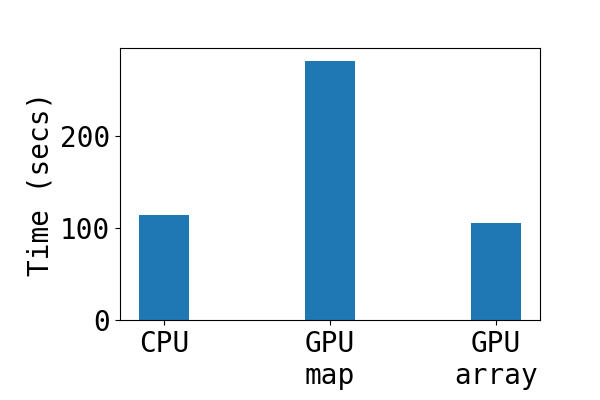}
			\label{fig:kosarak_grp_bottleneck}
		}
		\hfill
		\subfigure[DBLP, alternative C]
		{
			\includegraphics[width=.45\columnwidth]{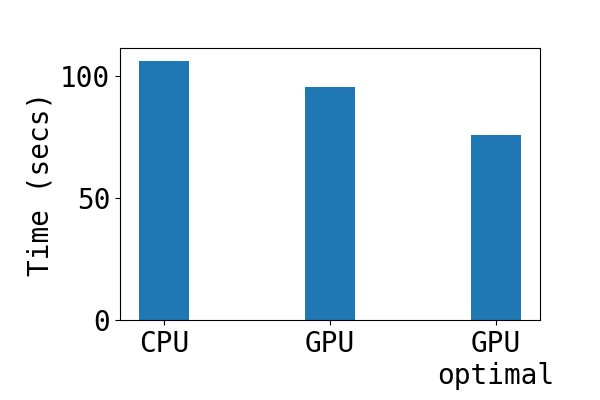}
			\label{fig:dblp_grp_bottleneck}
		}
		\caption{Comparison of GRP flavors ($t=0.5$, $M_{c}=4$GB)}
		\label{fig:grp_bottleneck}
	\end{figure}

	%The smallest performance speedup is measured for the GRP algorithm. As mentioned in Section~\ref{sec:impl}, we choose to split the workload and delegate a portion of candidates to the device.
	The KOSARAK dataset is more efficiently processed by GRP regardless of the threshold.
	In Figure~\ref{fig:kosarak_grp_bottleneck}, we compare the join times of the CPU and two GPU executions for the KOSARAK dataset. In the first GPU execution, we use a \textit{map} structure to delegate the whole verification phase to the device. In the second, we use raw arrays to delegate only the candidates generated in the first phase to the device, thus we split the verification workload between host and device. As it can be seen, the overhead imposed by using a map, which entails numerous memory checks, is not outweighed by faster verification. On the other hand, the drawback of assigning part of the verification to the CPU is that in datasets, such as KOSARAK, where the group expanding yields a larger number of candidates than the first phase, the host is assigned with significantly more verification workload than the device. Therefore, it is expected a GPGPU approach not to yield any benefits for such cases.

	Another issue during the expanding phase is that the host iterates and skips candidate pairs that are to be verified on the device. This adds extra overhead. On every dataset, except AOL, BMS and KOSARAK, the group expanding phase generates less candidates than the first phase. Nonetheless, we cannot avoid the candidate skipping overhead. In Figure~\ref{fig:dblp_grp_bottleneck}, we illustrate its impact on join time. For the DBLP dataset with $t=0.5$, the group expanding does not generate any candidate, thus, if we remove the expanding phase, we achieve a performance gain.
	
		\begin{figure*}[tb!]
		\centering
		\subfigure[Kosarak kernel execution]
		{
			\includegraphics[width=0.28\linewidth]{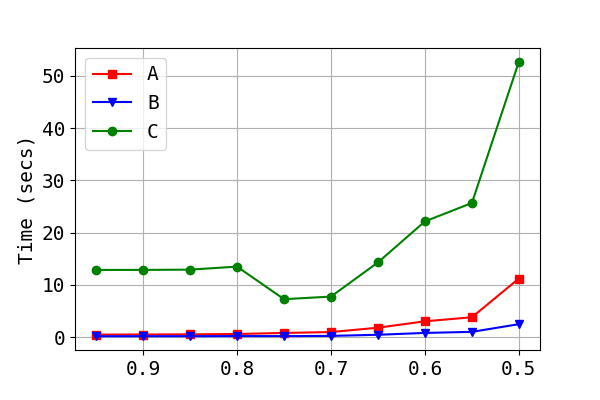}
			\label{fig:alt_comparison_kosarak_kernel}
		}
		\subfigure[Livejournal kernel execution]
		{
			\includegraphics[width=0.28\linewidth]{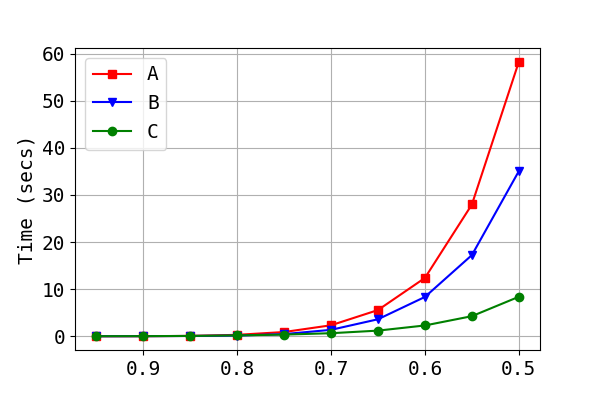}
			\label{fig:alt_comparison_livejournal_kernel}
		}
		\subfigure[Alternatives boundaries]
		{
			\includegraphics[width=0.25\linewidth]{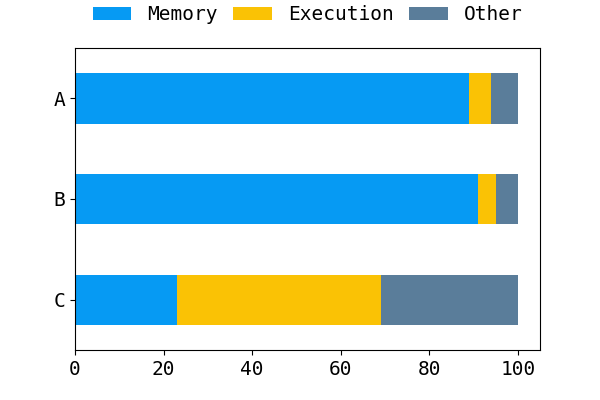}
			\label{fig:alt_comparison_profiling}
		}
		\caption{Alternatives comparison.}
		\label{fig:alt_comparison}
	\end{figure*}
	
	\subsection{Device Performance}
	
There are two kernel parameters, which require fine tuning depending on the dataset characteristics: (i) the verification alternative to be  followed and (ii) the thread block size.
%Furthermore, we conduct sensitivity analysis on specific experiments by profiling the device behavior.
	
	\subsubsection{Comparison of  our proposed alternatives }
	
	The two main differences between the verification alternatives is how threads access global memory and how they calculate the intersection of two sets. Each thread in alternative A accesses its respective candidate pair sets and calculates each intersection. Because of that, intra-warp divergence is maximized since a thread has its own execution path. Alternative B alleviates the performance because the threads of a block collaborate to load the corresponding probing set to shared memory and then, less candidate pairs per thread are verified. However, since each thread independently loads a candidate set and calculates the intersection, intra-warp divergence is still present. In alternative C, each block's threads collaborate first to load the probing and each candidate set to shared memory, and second, to perform the intersections. Therefore intra-warp divergence is low.
	
	For datasets with small average set size ($\le10$) such as AOL, BMS, KOSARAK, the global memory access footprint is also small, which renders alternatives A and B competitive. As shown in Figure~\ref{fig:alt_comparison_kosarak_kernel}, both A and B have similar performance for large thresholds, but for small ones alternative B performs better since thousands more thread blocks are launched and thus device occupancy is increased. On the contrary, alternative C seems infeasible for small set sizes, since the overhead to store them in shared memory dominates the verification time.
	The advantage of alternative C becomes apparent in datasets characterized by larger average set size such as DBLP, ENRON, LIVEJOURNAL and ORKUT. Figure~\ref{fig:alt_comparison_livejournal_kernel} shows the performance of the three alternatives for the LIVEJOURNAL dataset. As the number of candidate pairs increases for small thresholds, alternative C performs better since it achieves a higher warp execution efficiency (200\% and 33\% increase from A and B respectively). In addition, it has lower memory dependency as shown in Figure~\ref{fig:alt_comparison_profiling}. By minimizing the global memory access footprint, alternative C becomes mainly execution dependent. Other dependencies such as instruction fetch, instruction issue and synchronization arise, but they are less expensive than constantly accessing the global memory in a non-optimal way.
	
\textbf{Key result:} It is more beneficial to employ alternative B for sets of small size, and alternative C otherwise.

	\subsubsection{Block size}
	
	\begin{figure}[tb!]
		
		\subfigure[AOL, Alternative B]
		{
			\includegraphics[width=.45\columnwidth]{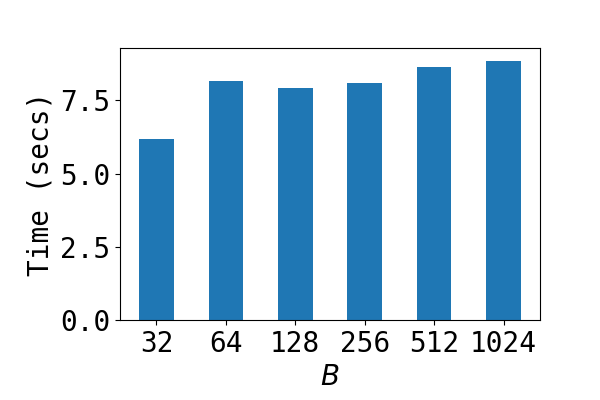}
			\label{fig:aol_block_size}
		}
		\hfill
		\subfigure[ORKUT, Alternative C]
		{
			\includegraphics[width=0.45\columnwidth]{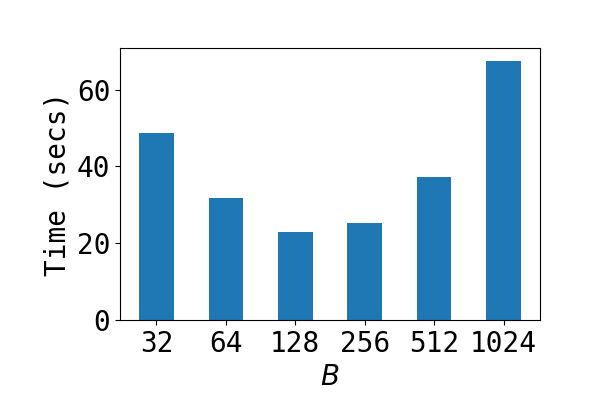}
			\label{fig:orkut_block_size}
		}
		\caption{Block size impact on verification time, $t=0.5$, $M_{c}=4GB$.}
		\label{fig:block_size}
	\end{figure}

	We investigate how the thread block size affects the verification time. For datasets with small average set size, where alternative B is preferred, $B=32$ has the best performance. Increasing $B$ in such datasets leads to a higher proportion of inactive warps and hence increases the verification time as shown in Figure~\ref{fig:aol_block_size}.	
	As the number of average set size in candidate pairs increases, alternative C is preferred to minimize the global memory access bottleneck. Alongside, fine tuning of $B$ is also required to achieve best device performance. For example, as shown in Figure~\ref{fig:orkut_block_size} for the ORKUT dataset, if we assign $B=32$, each thread receives more workload than optimal. By increasing $B$ up to 128, the workload is more evenly distributed since more threads contribute to the join. If we further increase $B$, this leads to a high number of inactive warps, and therefore to low warp execution efficiency.
	
\textbf{Key result:} Judiciously increasing $B$ when alternative C is employed  leads to higher performance. When combined with lower $M_c$, it manages to fully overlap verification and index/filtering phases.

	%\subsection{Discussion}
	
	%TODO
	%join time = index time stalls must verify each candidate pairs per iteration, with a small serialization overhead cost
	
	%our approach is tradeoff to introduce verification to a fast parallel environment, we believe in scale of billion set similarity as observed in the small thresholds for various datasets, exploit collaboration CPU-GPU
	
%	grid level alternative, not feasible explain, blocks cannot run concurrently in CUDA, for this alternative to be feasible a two-pass gpu invoke must be done per candidate pair

%	The same principles apply to the no-self join.

	\section{Related Work}
	\label{sec:rw}
	
	Although extensive research has been carried out on set similarity join for parallel paradigms, such as MapReduce\cite{VCL10,MF12,SHC14}, there are few studies investigating set similarity join on the GPGPU paradigm.
	An early proposal has appeared in \cite{lieberman:gpusjoin}, according to which Lieberman et al cast the similarity join operation as a GPU sort-and-search problem. First, they create a set of space-filling curves using bitonic sort on one of the input relations; then, they process each record from the other relation in parallel by executing searches in the space filling curves, using the Minkowski metric for similarity.
	In \cite{cruz2015gpu}, the authors employ the parallel-friendly MinHash algorithm to estimate the Jaccard similarity of two sets. Their solution is space-efficient since they only store set signatures instead of whole sets to perform the similarity join. However, due to the MinHash nature (i.e. data partitioning in \textit{bins}), fine-tuning is required to achieve balance between accuracy (to avoid false positives) and execution time. The main limitations of the above techniques is that they are approximate, whereas we propose solutions to the problem of exact set similarity joins, and we are not inherently limited to Jaccard similarity only.
	
	The only known work to date on exact similaity joisn on GPUs has appeared in \cite{Ribeiro-JuniorQ17}. This work allocates both the indexing/filtering and the verification phase on GPUs and exhibits promising results. As an extension of our work, we also aim to investigate design alternatives so that both set similarity join phases are performed on GPUs.
	
	Similarity joins  are also discussed in \cite{bohm2009indexsupported}, where two nested loop join (NLJ) algorithms are presented: a naive NLJ and a faster index-supported NLJ. The index is created on the CPU side during the preprocessing phase. Both algorithms use the Euclidean distance for similarity and thus they are not suitable for set similarity joins. Nevertheless, our solutions also perform sophisticated CPU-side indexing before the GPU-side processing.
	
	Another problem, which is close to set similarity join and has been studied on the GPGPU paradigm, is similarity (nearest neighbor) search. Examples include \cite{pan2011fast,zhou2016generic,wang2017flash,Wieschollek_2016_CVPR,johnson2017billion} but none of them can be applied to our problem

%The proposal in \cite{pan2011fast,zhou2016generic} uses a GPU-based parallel Locality Sensitive Hashing (LSH) algorithm to perform approximate $k$-nearest neighbor ($k$NN) search. In \cite{wang2017flash}, a hybrid CPU-GPU framework, which uses LSH combined with other techniques, such as reservoir sampling is presented, where, the CPU constructs hashtables and the GPU process them and performs a count-based top-$k$ selection. In \cite{Wieschollek_2016_CVPR}, the authors propose a two-level quantization tree and a re-ranking method for fast approximate nearest  neighbor search. In \cite{johnson2017billion}, Johnson et al. present an optimal designed framework to compute a $k$-NN graph.

	\section{Conclusion}
	
	\label{sec:concl}
	
This work describes the first thorough ivestigation to date regarding the design alternatives for the verification phase in exact set similarity joins using a GPU. We conform to the established filter-verification framework, and we transfer verification to the GPU. We provide solutions to the issues involved, such as data layout on the device memory, serialization, and thread workload. Using real datasets, we show that we manage to fully overlap the GPU tasks with the CPU ones, when the candidate pairs are at the order of tens of billions of sets.
Despite the significant speed-ups reported, in several cases, exact set similarity remains a very expensive task due to the non-parallelized filter phase. The main direction for future work is to explore GPU-tailored solutions for filtering as well in line with works such as \cite{Ribeiro-JuniorQ17,zhou2016generic}, building upon also recent advances for CPUs, such as those in
 \cite{wang2017leveraging}.

	% or
	

	\ifCLASSOPTIONcompsoc
	\section*{Acknowledgments}
	\else
	\section*{Acknowledgment}
	\fi
	
	The authors gratefully acknowledge the support of NVIDIA Corporation through the donation of the GPU used.
	
	\ifCLASSOPTIONcaptionsoff
	\newpage
	\fi
	
\bibliographystyle{abbrv}
\bibliography{literature}

\begin{thebibliography}{10}

\bibitem{ashkiani2017dynamic}
S.~Ashkiani, M.~Farach-Colton, and J.~D. Owens.
\newblock A dynamic hash table for the gpu.
\newblock {\em arXiv preprint arXiv:1710.11246}, 2017.

\bibitem{BML10}
R.~Baraglia, G.~D.~F. Morales, and C.~Lucchese.
\newblock Document similarity self-join with mapreduce.
\newblock In {\em ICDM}, pages 731--736, 2010.

\bibitem{BayardoMS07}
R.~J. Bayardo, Y.~Ma, and R.~Srikant.
\newblock Scaling up all pairs similarity search.
\newblock In {\em Proceedings of the 16th International Conference on World
  Wide Web, {WWW} 2007, Banff, Alberta, Canada, May 8-12, 2007}, pages
  131--140, 2007.

\bibitem{BellasG17}
C.~Bellas and A.~Gounaris.
\newblock {GPU} processing of theta-joins.
\newblock {\em Concurrency and Computation: Practice and Experience}, 29(18),
  2017.

\bibitem{bohm2009indexsupported}
C.~B{\"o}hm, R.~Noll, C.~Plant, and A.~Zherdin.
\newblock Index-supported similarity join on graphics processors.
\newblock In {\em BTW}, volume 144, pages 57--66, 2009.

\bibitem{BourosGM12}
P.~Bouros, S.~Ge, and N.~Mamoulis.
\newblock Spatio-textual similarity joins.
\newblock {\em {PVLDB}}, 6(1):1--12, 2012.

\bibitem{cheng2014professional}
J.~Cheng, M.~Grossman, and T.~McKercher.
\newblock {\em Professional Cuda C Programming}.
\newblock John Wiley \& Sons, 2014.

\bibitem{cruz2015gpu}
M.~S. Cruz, Y.~Kozawa, T.~Amagasa, and H.~Kitagawa.
\newblock Gpu acceleration of set similarity joins.
\newblock In {\em International Conference on Database and Expert Systems
  Applications}, pages 384--398, 2015.

\bibitem{green2012gpu}
O.~Green, R.~McColl, and D.~A. Bader.
\newblock Gpu merge path: a gpu merging algorithm.
\newblock In {\em Proceedings of the 26th ACM international conference on
  Supercomputing}, pages 331--340, 2012.

\bibitem{green2014fast}
O.~Green, P.~Yalamanchili, and L.-M. Mungu{\'\i}a.
\newblock Fast triangle counting on the gpu.
\newblock In {\em Proceedings of the 4th Workshop on Irregular Applications:
  Architectures and Algorithms}, pages 1--8, 2014.

\bibitem{JLFL14}
Y.~Jiang, G.~Li, J.~Feng, and W.~Li.
\newblock String similarity joins: An experimental evaluation.
\newblock {\em {PVLDB}}, 7(8):625--636, 2014.

\bibitem{johnson2017billion}
J.~Johnson, M.~Douze, and H.~J{\'e}gou.
\newblock Billion-scale similarity search with gpus.
\newblock {\em arXiv preprint arXiv:1702.08734}, 2017.

\bibitem{KDK+11}
S.~W. Keckler, W.~J. Dally, B.~Khailany, M.~Garland, and D.~Glasco.
\newblock Gpus and the future of parallel computing.
\newblock {\em IEEE Micro}, 31(5):7--17, 2011.

\bibitem{cuda-book}
D.~B. Kirk and W.~W. Hwu.
\newblock {\em Programming Massively Parallel Processors - {A} Hands-on
  Approach, 2nd Ed.}
\newblock Morgan Kaufmann, 2013.

\bibitem{lieberman:gpusjoin}
M.~D. Lieberman, J.~Sankaranarayanan, and H.~Samet.
\newblock A fast similarity join algorithm using graphics processing units.
\newblock In {\em Data Engineering, 2008. ICDE 2008. IEEE 24th International
  Conference on}, pages 1111--1120. IEEE, 2008.

\bibitem{pel}
W.~Mann and N.~Augsten.
\newblock Pel: Position-enhanced length filter for set similarity joins.
\newblock In {\em Proceedings of the 26th GI-Workshop Grundlagen von
  Datenbanken}, pages 89--94, 2014.

\bibitem{mann:ssjoin}
W.~Mann, N.~Augsten, and P.~Bouros.
\newblock An empirical evaluation of set similarity join techniques.
\newblock {\em Proceedings of the VLDB Endowment}, 9(9):636--647, 2016.

\bibitem{MF12}
A.~Metwally and C.~Faloutsos.
\newblock V-smart-join: A scalable mapreduce framework for all-pair similarity
  joins of multisets and vectors.
\newblock {\em PVLDB}, 5(8):704--715, 2012.

\bibitem{mittal2015survey}
S.~Mittal and J.~S. Vetter.
\newblock A survey of cpu-gpu heterogeneous computing techniques.
\newblock {\em ACM Computing Surveys (CSUR)}, 47(4):69, 2015.

\bibitem{pan2011fast}
J.~Pan and D.~Manocha.
\newblock Fast gpu-based locality sensitive hashing for k-nearest neighbor
  computation.
\newblock In {\em Proceedings of the 19th ACM SIGSPATIAL international
  conference on advances in geographic information systems}, pages 211--220.
  ACM, 2011.

\bibitem{RIBEIRO201162}
L.~A. Ribeiro and T.~Härder.
\newblock Generalizing prefix filtering to improve set similarity joins.
\newblock {\em Information Systems}, 36(1):62 -- 78, 2011.

\bibitem{Ribeiro-JuniorQ17}
S.~Ribeiro{-}J{\'{u}}nior, R.~D. Quirino, L.~A. Ribeiro, and W.~S. Martins.
\newblock Fast parallel set similarity joins on many-core architectures.
\newblock {\em {JIDM}}, 8(3):255--270, 2017.

\bibitem{SHC14}
A.~D. Sarma, Y.~He, and S.~Chaudhuri.
\newblock Clusterjoin: {A} similarity joins framework using map-reduce.
\newblock {\em {PVLDB}}, 7(12):1059--1070, 2014.

\bibitem{VCL10}
R.~Vernica, M.~J. Carey, and C.~Li.
\newblock Efficient parallel set-similarity joins using mapreduce.
\newblock In {\em SIGMOD Conference}, pages 495--506, 2010.

\bibitem{WangLF12}
J.~Wang, G.~Li, and J.~Feng.
\newblock Can we beat the prefix filtering?: an adaptive framework for
  similarity join and search.
\newblock In {\em Proc. of {ACM} {SIGMOD}}, pages 85--96, 2012.

\bibitem{wang2017leveraging}
X.~Wang, L.~Qin, X.~Lin, Y.~Zhang, and L.~Chang.
\newblock Leveraging set relations in exact set similarity join.
\newblock {\em Proceedings of the VLDB Endowment}, 10(9):925--936, 2017.

\bibitem{wang2017flash}
Y.~Wang, A.~Shrivastava, and J.~Ryu.
\newblock Flash: Randomized algorithms accelerated over cpu-gpu for ultra-high
  dimensional similarity search.
\newblock {\em arXiv preprint arXiv:1709.01190}, 2017.

\bibitem{Wieschollek_2016_CVPR}
P.~Wieschollek, O.~Wang, A.~Sorkine-Hornung, and H.~P.~A. Lensch.
\newblock Efficient large-scale approximate nearest neighbor search on the gpu.
\newblock In {\em The IEEE Conference on Computer Vision and Pattern
  Recognition (CVPR)}, June 2016.

\bibitem{XiaoWLYW11}
C.~Xiao, W.~Wang, X.~Lin, J.~X. Yu, and G.~Wang.
\newblock Efficient similarity joins for near-duplicate detection.
\newblock {\em {ACM} Trans. Database Syst.}, 36(3):15:1--15:41, 2011.

\bibitem{zhou2016generic}
J.~Zhou, Q.~Guo, H.~Jagadish, W.~Luan, A.~K. Tung, Y.~Yang, and Y.~Zheng.
\newblock Generic inverted index on the gpu.
\newblock {\em arXiv preprint arXiv:1603.08390}, 2016.

\end{thebibliography}
	
\end{document}